\documentclass[journal,10pt]{IEEEtran}
\usepackage{subfigure}
\PassOptionsToPackage{subfigure}{tocloft}
\usepackage{CJK}
\usepackage{algorithm}
\usepackage{algorithmic}
\usepackage{psfrag}
\usepackage{graphicx}
\usepackage{epstopdf}
\usepackage{multirow}
\usepackage{stfloats}
\usepackage{cite}
\usepackage{arydshln}
\usepackage{enumerate}
\usepackage{bbm}
\usepackage{bm}
\usepackage{extarrows}

\usepackage{multirow}
\usepackage[dvipsnames]{xcolor}

\usepackage{amsthm}
\usepackage{amsmath}
\usepackage{amssymb}

\theoremstyle{plain}

\newtheorem{lemma}{{Lemma}}

\usepackage{xcolor}
\usepackage{changes}

\usepackage{lipsum}

\title{Cascaded Channel Estimation for \\Intelligent Reflecting Surface Assisted\\ Multiuser MISO Systems}

\author{
Huayan Guo, \IEEEmembership{Member, IEEE},  and Vincent K. N. Lau,  \IEEEmembership{Fellow, IEEE}

\thanks{
The work of H. Guo and V. K. N. Lau  was supported by the Hong Kong Research Grants Council Collaborative Research Fund 2020/21 Collaborative Research Project Grant (CRPG) C6012-20GF.
}
\thanks{The authors are with the Department of Electronics and Computer Engineering, The Hong Kong University of Science and Technology, Hong Kong 999077, China (e-mail: \{eeguohuayan, eeknlau\}@ust.hk).}
}

\begin{document}

\maketitle

\begin{abstract}
This paper investigates the  uplink cascaded channel estimation for intelligent-reflecting-surface (IRS)-assisted multi-user multiple-input-single-output systems. We focus on a sub-6 GHz scenario where the channel propagation is not sparse and  the number of IRS elements can be larger than the number of BS antennas. A novel channel estimation protocol without the need of on-off amplitude control to avoid the reflection power loss is proposed. In addition, the pilot overhead is substantially reduced by exploiting the common-link structure to decompose the cascaded channel coefficients by the multiplication of the common-link variables and the user-specific variables. However, these two types of variables are highly coupled, which makes them difficult to estimate. To address this issue, we formulate an optimization-based joint channel estimation problem, which only utilizes the covariance of the cascaded channel. Then, we design a low-complexity alternating optimization algorithm with efficient initialization for the non-convex optimization problem, which achieves a local optimum solution. To further enhance the estimation accuracy, we propose a new formulation to optimize the training phase shifting configuration for the proposed protocol, and then solve it using the successive convex approximation algorithm. Comprehensive simulations verify that the proposed algorithm has supreme performance compared to various state-of-the-art baseline schemes.
\end{abstract}


\section{Introduction}


Intelligent reflecting surface (IRS) is an artificial planar array consisting of numerous reconfigurable passive elements with the capability of manipulating the impinging electromagnetic signals and offering anomalous reflections \cite{Tan2018SRA,cuiTJ2017metasurface,Larsson2020Twocritical,Garcia2020JSACIRS_gap_scatter_reflect}.
Many recent studies have indicated that IRS is a promising solution to build a programmable wireless environment via steering the incident signal in fully customizable ways  to enhance the spectral and energy efficiency of legacy systems \cite{RenzoJSACposition,magzineWuqq,Liaskos2018magzineIRS,Renzo2019position}.
Most contributions in this area focus on joint active and passive procoding design with various objectives and constraints \cite{Wuq2019TWCprecoderIRS,HY2020TWCprecoderIRS,mux2020TWCprecoderIRS,ZhouG2020TSPIRSprecoderimperfectCE,LinS2021TWCprecoderIRS}.
The potential gains claimed by these works highly depend on the availability of accurate channel state information (CSI).
However, channel estimation is a challenging task for the IRS-assisted system because there are no sensing elements or radio frequency  chains, and thus there is no baseband processing capability in the IRS.

\begin{figure}
[!t]
\centering
\includegraphics[width=.8\columnwidth]{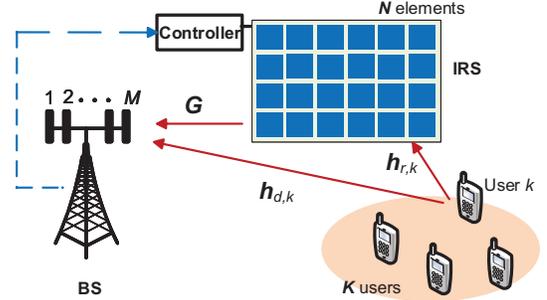}
\caption{The IRS-assisted  multiuser MISO communication system.}
\label{IRS_system}
\end{figure}

\begin{figure}
[!t]
  \centering
  \subfigure[Protocol for SU-MISO system]{
    \label{protocol:a} 
    \includegraphics[width=.6\columnwidth]{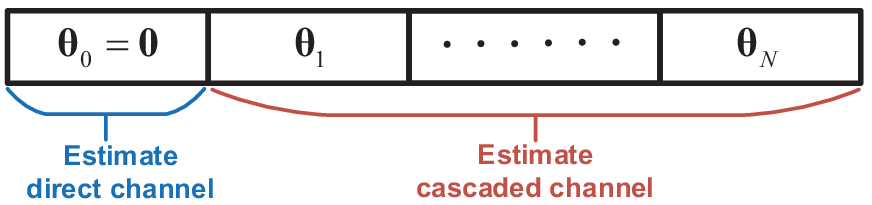}}
  \subfigure[Directly extended protocol for MU-MISO system]{
    \label{protocol:b} 
    \includegraphics[width=.6\columnwidth]{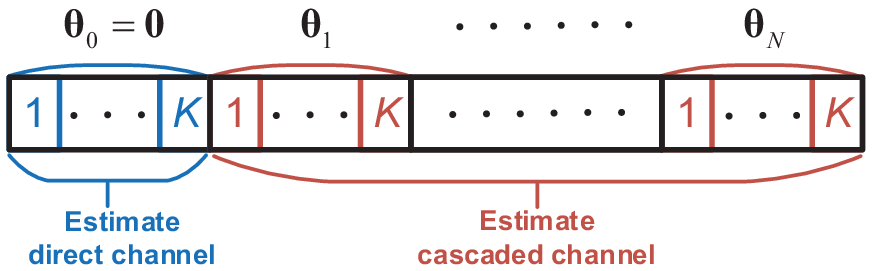}}
  \caption{The conventional uplink channel estimation protocols without exploiting the common-link structure.}
  \label{protocol_tran} 
\end{figure}

Some early-attempted works \cite{Jensen2020SUCE,zhangruiJSACSUCE,YouxhSPLCESU} estimate the uplink cascaded IRS channel  for single user (SU) multiple-input-single-output (MISO) systems using the protocol shown in {\figurename~\ref{protocol:a}}.
In these works, the cascaded channel is equivalently represented as a traditional $M\times N$ MIMO channel, where $M$ and $N$ is the base station (BS) array size and the IRS size, respectively, and the sensing matrix for channel reconstruction consists of the phase shifting vectors in consecutive training timeslots.
Many works \cite{Araujo2021JSAC_CE_PARAFAC,Mishra2019CEonoff,Elbir2020WCL_DL_CE,
Zhouzy2020decompositionCE,Kundu2021OJCSLMMSE_DFTGOOD,Alwazani2020OJCSLMMSE_DFT} directly extend the SU protocol to multi-user (MU) MISO systems, in which $K$ users transmit orthogonal pilot sequences in each training timeslot, as shown in {\figurename~\ref{protocol:b}}.
Based on this protocol,
the on-off IRS state (amplitude) control strategy is proposed in \cite{Araujo2021JSAC_CE_PARAFAC,Mishra2019CEonoff,Elbir2020WCL_DL_CE} to better decompose the MU cascaded channel coefficients   for easier channel estimation of the cascaded channel for each user.

It is pointed out by \cite{Liuliang_CE2020TWC} that direct application of the SU protocol on an MU-MISO system fails to exploit the structural property and results in substantially larger pilot overheads.
Intuitively, all the cascaded channels share a common BS-IRS link and it is possible to reduce the pilot overhead since the number of independent variables is $MN+NK$ instead of $MNK$.
One algorithm is proposed in \cite{DaiLLFullD} with the idea of sequentially estimating the BS-IRS channel and the IRS-user channels.
However, it requires that the BS can work at full-duplex mode.
For the cascaded channel estimation,
a new channel estimation protocol is proposed in \cite{Liuliang_CE2020TWC}.
Specifically, the cascaded channel of one reference user is firstly estimated based on the SU protocol, and then other users' channels are estimated by only estimating the ratios of their channel coefficients to the reference channel, which can be referred to as the relative channels.
The overall training overhead is reduced from $NK$ to $K+N+\lceil \frac{N}{M}\rceil(K-1)$.
However, there is an error propagation issue associated with this scheme  since a low-accuracy estimation on the reference  channel may jeopardize  the estimations of the relative channels.
Moreover, some IRS elements need to be switched off while estimating the relative channels for coefficients decomposition \cite{Liuliang_CE2020TWC}.

For IRS design, the ``off'' state means no reflection (i.e., perfectly absorbing the incident signals),  and hence  it is difficult \cite{Perfect_Absorption1,Perfect_Absorption2,Perfect_Absorption3} and also attracts additional implementation costs since this state is unnecessary for data transmission after the channel estimation.
In addition, switching off the IRS elements causes reflection power loss, which will lower the receive signal-to-noise ratio (SNR).
Some recent works attempt to overcome this issue using ``always-ON'' training schemes.
In \cite{Zhouzy2020decompositionCE,Kundu2021OJCSLMMSE_DFTGOOD,Alwazani2020OJCSLMMSE_DFT}, cascaded channel estimation algorithms based on tensor decomposition  are proposed for MU-MISO systems, without requiring selected IRS elements to be off using the protocol in {\figurename~\ref{protocol:b}}.
In particular, the training phase shifts are optimized to minimize the mean squared error (MSE), and it has been verified that the discrete-Fourier-transform (DFT)-based training phase shifting configuration is optimal in this scenario.
However, the pilot overhead is $NK$ since the protocol in {\figurename~\ref{protocol:b}} does not utilize the
common-link property.
In \cite{double_IRS}, an always-ON training scheme is proposed, which extends the protocol in \cite{Liuliang_CE2020TWC} to the double-IRS aided system.
However, the number of BS antennas needs to be equal to or larger than the number of IRS elements (i.e., $M\geq N$) to guarantee a full-rank measurement matrix to estimate the relative channels between the reference user and the other users.\footnote{
According to the property ${\rm{rank}}({\bf A} \otimes{\bf B})={\rm{rank}}({\bf A} ){\rm{rank}}({\bf B} )$, the rank of the measurement matrix in equation (40) of \cite{double_IRS} cannot be larger than $(K-1)M$ while the targeted rank is $(K-1)N$.
}
This assumption is quite restrictive as the number of elements of the IRS ($N$) is usually larger than the number of antennas at the BS.


Another critical problem is the feasibility issue when the channel statistical prior information is utilized to improve the channel estimation accuracy,
although this is a common idea for conventional MIMO channel estimation.
In \cite{Kundu2021OJCSLMMSE_DFTGOOD} and \cite{Alwazani2020OJCSLMMSE_DFT}, statistical knowledge of the individual BS-IRS link and IRS-user links is required. However, these messages are not available in practice since
none of the existing algorithms, to the best of the authors' knowledge, can reconstruct the individual channel coefficients when $M<N$ due to the ambiguity issue, which introduces a random scaling on the estimated individual channels \cite{Araujo2021JSAC_CE_PARAFAC,Hezqwcl2020IRS_CE_SU}.
In \cite{Liuliang_CE2020TWC}, the linear minimum mean squared error (LMMSE) estimator is adopted to estimate the relative channels. However, the LMMSE estimator requires the covariance of the relative channels, which is also difficult since in general the distribution of the relative channels is heavy-tailed.




In this paper, we focus on the uplink cascaded IRS channel estimation for an MU-MISO system in sub-6 GHz bands.
In these frequency bands, the IRS design will be easier, but the channel propagation is not sparse.\footnote{If the IRS operates at millimeter wave frequencies, some sparse channel estimation algorithms have been further proposed based on the protocol in {\figurename~\ref{protocol:b}} with less training timeslots   by exploiting the potential channel sparsity \cite{Hezqwcl2020IRS_CE_SU,DaiLL2021CLpartII,Wangpl2020splCS_CE,Ardah2021splCS_TRICE,YuanXJ2020JSAConoffsparseCE,LClearning}.
}
In addition, the number of IRS elements in the system can be larger than the number of BS antennas (i.e., $N>M$).
We propose a holistic solution to address the aforementioned issues.
In particular, a novel always-ON training protocol is designed; meanwhile the common-link structure is utilized to reduce the pilot overhead.
Furthermore, an optimization-based cascaded channel estimation framework, which is flexible to utilize more practical channel statistical prior information, is proposed.
The following summarizes our key contributions.
\begin{itemize}
\item  {\bf Always-ON Channel Estimation Protocol Exploiting the Common-Link Structure}:
    We propose a novel channel estimation protocol without the need for on-off amplitude control to avoid the reflection power loss.
    Meanwhile, the common-link structure is exploited and the pilot overhead is reduced to $K+N+\lceil \frac{N}{M}\rceil(K-1)$.

    In addition, the proposed protocol is applicable with any number of  elements at the IRS (also including $N \leq M$).
    Further, it does not need a ``reference user'', and as such, the estimation performance is enhanced owing to the multiuser diversity.


\item  {\bf Optimization-Based Cascaded Channel Estimation Framework}:
    Since there is no on-off amplitude control, the cascaded channel coefficients are highly coupled.
    In order to exploit the  common-link structure, we decompose the cascaded channel coefficients by the multiplication of the common-link variables and the user-specific variables, and then an optimization-based joint channel estimation problem is formulated based on the maximal a posterior probability (MAP) rule. The proposed optimization-based approach is flexible to incorporate different kinds of channel statistical prior setups. Specifically, we utilize the combined statistical information of the cascaded channels, which is a weaker requirement compared to statistical knowledge of the individual BS-IRS and IRS-user channels.
    Then, a low-complexity alternating optimization algorithm is  proposed to achieve a local optimum solution.
    Simulation results demonstrated that the optimization solution with proposed protocol achieves a more than $15$ dB gain compared to the benchmark.

\item  {\bf Training Phase Shifting Optimization for the Proposed Protocol}:
    The phase shifting configuration can substantially enhance the channel estimation performance of the cascaded IRS channel because the phase shifting vectors are important components in the measurement matrix for channel reconstruction.
    However, traditional solutions \cite{Zhouzy2020decompositionCE,Kundu2021OJCSLMMSE_DFTGOOD,Alwazani2020OJCSLMMSE_DFT} of phase shifting optimization for SU cascaded IRS channel estimation cannot be directly applied to the MU case due to that  the cascaded channel coefficients are highly coupled when the common-link structure is exploited.
    We propose a new formulation to optimize the phase shifting configuration, which maximizes the average reflection gain of the IRS.
    Simulation results further verify the proposed configuration achieves a more than $3$ dB gain compared to the state-of-the-art  baselines.
\end{itemize}

\section{System Model}\label{system model}

\subsection{System Model of MU-MISO IRS Systems}
This paper investigates the uplink channel estimation in a narrow-band IRS-aided MU-MISO communication system that consists of one BS with $M$ antennas, one IRS with $N$ elements, and $K$ single-antenna users,\footnote{
{We adopt the single-antenna-user setup here for ease of presentation. The signal model can be directly extended to the setup when users have multiple antennas by transmitting orthogonal uplink pilot sequences in different antennas.
}}
as illustrated in {\figurename~\ref{IRS_system}}.
Let ${\bf h}_{{\rm d},k} \in {\mathbb C}^{M \times 1}$ denote the BS-user channel (a.k.a., the direct channel) for user $k$, ${\bf G} \in {\mathbb C}^{N \times M}$ denote the common BS-IRS channel, and ${\bf h}_{{\rm r},k} \in {\mathbb C}^{N \times 1}$ denote the IRS-user channel for user $k$.
We assume  quasi-static block fading  for all the channels such that the channel coefficients remain constant within one channel coherence interval, and they are independent and identically distributed (i.i.d.) between coherence intervals.  Note that the quasi-static model considers the worse case scenarios where the temporal correlations between blocks are not exploited. In practice, the pilot overheads can be further reduced if one exploits temporal correlations of the channel blocks \cite{kalman_filter_IRS2021TVT,kalman_filter_IRS2021chinacom}, but this is outside the scope of the paper.

The received baseband signal at the BS is given by
\begin{equation}\label{equ:y_model1}
\begin{aligned}[b]
{\bf y}_{t}&=\sum_{k=1}^K \left({\bf h}_{{\rm d},k}+ {\bf G}^{\rm T} {\bf \Theta}_t  {\bf h}_{{\rm r},k} \right) x_{k,t}
+{ {\bf z}_t}
,
\end{aligned}
\end{equation}
where $t$ is the time index, $x_{k,t}$ is the transmit pilot symbol from user $k$, ${\bf z}_t \sim {\cal{CN}}({\bm 0},\sigma_0^2 {\bf I}_{M}) $ is the {additive white Gaussian noise} (AWGN), and ${\bf \Theta}_t \in {\mathbb C}^{N \times N}$ is the IRS reflection coefficient matrix.
It is known that ${\bf \Theta}_t$ is a diagonal matrix such that ${\bf \Theta}_t={\rm diag}({\bm \theta}_t)$, where ${\bm \theta}_t=[e^{\jmath \varphi_{t,1}},e^{\jmath \varphi_{t,2}},\cdots,e^{\jmath \varphi_{t,N}}]^{\rm T}$ is the phase shifting vector from the IRS.\footnote{
{In practice, the reflection efficiency cannot be 1, which is known as the reflection loss. However, for ease of presentation, this loss can be absorbed into the path loss of ${\bf G}$ since it is a constant value.
}}

\subsection{IRS Cascaded Channel Model}\label{sec:model_cascaded}

Denote the cascaded channel related to the BS's $m$-th antenna and the $k$-th user by
\begin{equation}\label{equ:cascaded_channel_vector}
\begin{aligned}[b]
{\bf h}_{{\rm I},k,m}={\rm{diag}}({\bf h}_{{\rm r},k}) {\bf g}_m,
\end{aligned}
\end{equation}
where ${\bf g}_m$ is the $m$-th column in ${\bf G}=[{\bf g}_1,\cdots,{\bf g}_M]$.
The cascaded channel over all BS antennas is given by ${\bf H}_{{\rm I},k}=[{\bf h}_{{\rm I},k,1}^{\rm T},\cdots,{\bf h}_{{\rm I},k,M}^{\rm T}]^{\rm T}$, and we have
\begin{equation}\label{equ:cascaded_channel}
\begin{aligned}[b]
{\bf H}_{{\rm I},k}={\bf G}^{\rm T} {\rm{diag}}({\bf h}_{{\rm r},k}).
\end{aligned}
\end{equation}
Substituting \eqref{equ:cascaded_channel} into \eqref{equ:y_model1}, the received signal is given by
\begin{equation}\label{equ:y_model_vn}
\begin{aligned}[b]
{\bf y}_t&=\sum_{k=1}^K \left({\bf h}_{{\rm d},k}+ {\bf H}_{{\rm I},k} {\bm{\theta}}_t \right) x_{k,t}
+{\bf z}_t
.
\end{aligned}
\end{equation}

In \cite{Araujo2021JSAC_CE_PARAFAC,Mishra2019CEonoff,Elbir2020WCL_DL_CE} and
\cite{Zhouzy2020decompositionCE,Kundu2021OJCSLMMSE_DFTGOOD,Alwazani2020OJCSLMMSE_DFT}, ${\bf H}_{{\rm I},k}$ for all $k$  are estimated
without exploiting the implicit common link structure behind the ${\bf H}_{{\rm I},k}$ for all $k$. As a result, there are $MNK$ variables to be estimated and this poses a heavy penalty on the required pilot overheads in the MU-MISO system.
On the other hand, from \eqref{equ:cascaded_channel}, we can see that the cascaded channels $\left\{{\bf H}_{{\rm I},1}, {\bf H}_{{\rm I},2},\cdots,{\bf H}_{{\rm I},K}\right\}$ all share a common BS-IRS link $\bf G$.
Specifically,
 ${\bf H}_{{\rm I},k}$ is the multiplication of the common $\bf G$ and the user-specific ${\rm{diag}}({\bf h}_{{\rm r},k})$.
In other words, the cascaded channels are not independent variables. In fact,  the common link structure $\bf G$ should be exploited in the channel estimation.
As a result, the total number of independent variables
is reduced to $MN+NK$.

We assume ${\mathbb E}\left[ {\bf h}_{{\rm I},k,m}\right]={\bf 0}$ for all $k$ and $m$.\footnote{
{The proposed algorithm in this paper is applicable for the case when an LoS link exists and ${\mathbb E}\left[ {\bf h}_{{\rm I},k,m}\right]\neq{\bf 0}$. Simply substitute ${\mathbb E}\left[ {\bf h}_{{\rm I},k,m}\right]$ into the prior distribution model in \eqref{equ:prior}. Note that since the rank-1 LoS link usually is very strong and easier to be estimated, it will dominate the power of the channel coefficients, and the NMSE will be better than the NLoS scenario investigated in this paper.
}}
The covariance of the cascaded channel ${\bf h}_{{\rm I},k,m}$ is given by
\begin{equation}\label{equ:cascaded_channel_statisical}
\begin{aligned}[b]
{\bf C}_m^{(k)}={\mathbb E}\left[ {\bf h}_{{\rm I},k,m} {\bf h}_{{\rm I},k,m}^{\rm H} \right].
\end{aligned}
\end{equation}
In this paper, we focus on the case when ${\bf C}_m^{(k)}$ is a full rank matrix for all $m$ and $k$, which is generally true in sub-6 GHz bands.
We  design a channel estimation algorithm and phase shifting configuration scheme by utilizing knowledge of  ${\bf C}_m^{(k)}$.
Note that in \cite{Kundu2021OJCSLMMSE_DFTGOOD} and \cite{Alwazani2020OJCSLMMSE_DFT}, the channel estimation algorithms for the cascaded channels require knowledge of the covariance of the IRS-BS link $\bf G$ as well as the covariance of the IRS-user links ${\bf h}_{{\rm r},k}$. Also note that knowledge of the covariance of the cascaded channel ${\bf C}_m^{(k)}$ is a weaker requirement compared to knowledge of the individual covariances $\bf G$ and ${\bf h}_{{\rm r},k}$.



\section{Proposed Channel Estimation Protocol}\label{dense_scheme}


\subsection{Overview of the Selected On-Off Channel Estimation Protocol}\label{overview_onoff}
\begin{figure}
[!ht]
\centering
\includegraphics[width=.95\columnwidth]{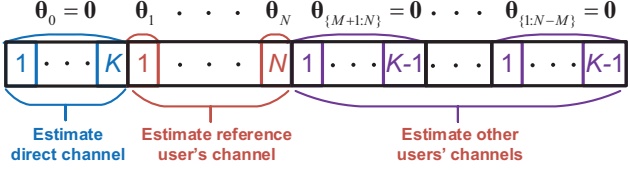}
\caption{Selected on-off channel estimation protocol in \cite{Liuliang_CE2020TWC}.}
\label{protocol:c}
\end{figure}

The selected on-off channel estimation protocol in \cite{Liuliang_CE2020TWC} is illustrated in {\figurename~\ref{protocol:c}}, and consists of three stages.
In stage I, the BS-user channels are estimated by switching off all the IRS elements.
In stage II, a reference user is selected, which is indexed by user $1$, and its cascaded channel ${\bf H}_{{\rm I},1}$ is estimated using the algorithm for SU-MISO cases \cite{zhangruiJSACSUCE}.
In stage III, the other $K-1$ users' cascaded channels are estimated by exploiting the common-link property.

We focus on the estimation in stage III. Substituting ${\bf H}_{{\rm I},1}={\bf G}^{\rm T} {\rm{diag}}({\bf h}_{{\rm r},1})$ into \eqref{equ:y_model1}, the received signal is given by
\begin{equation}\label{equ:y_model_LL}
\begin{aligned}[b]
{\bf y}_t&=\sum_{k=1}^K {\bf h}_{{\rm d},k} x_{k,t}+
{\bf H}_{{\rm I},1} {\bm{\theta}}_t x_{1,t}\\
&\qquad+\sum_{k=2}^K {\bf H}_{{\rm I},1} {\rm diag}({\bm \theta}_t)  {\bf h}_{{\rm u},k}  x_{k,t}
+{ {\bf z}_t}
,
\end{aligned}
\end{equation}
where ${\bf h}_{{\rm u},k} ={\rm diag}({\bf h}_{{\rm r},1})^{-1}{\bf h}_{{\rm r},k}$ for all $k=2,3,\cdots,K$, which are the user-specific variables to be estimated in this stage after exploiting ${\bf H}_{{\rm I},1}$ as the common-link variable.
In \cite{Liuliang_CE2020TWC}, to estimate ${\bf h}_{{\rm u},k}$, only the $k$-th user sends $x_{k,t}=1$ and all the other users are inactive such that $x_{j,t}=0$ for all $j \neq k$.
The received signal is given by
\begin{equation}
\begin{aligned}[b]
{\bf y}_t
&={\bf h}_{{\rm d},k}+ {\bf H}_{{{\rm I},1}} {\rm diag}({\bm \theta}_t)  {\bf h}_{{\rm u},k}
+ {\bf z}_t
.
\end{aligned}
\end{equation}
By only switching on the first $M$ IRS elements with $[\theta_{t,1},\theta_{t,2},\cdots,\theta_{t,M}]^{\rm T}={\bf 1}$, the first $M$ coefficients in ${\bf h}_{{\rm u},k}$ can be estimated by
\begin{equation}
\begin{aligned}[b]
\begin{bmatrix}{h}_{{\rm u},k,1}\\ \vdots \\ { h}_{{\rm u},k,M}\end{bmatrix}
=
{
\begin{bmatrix}
    H_{{{\rm I},1},11} & \dots & H_{{{\rm I},1},1M}\\
    \vdots  & \ddots & \vdots\\
    H_{{{\rm I},1},M1} & \dots & H_{{{\rm I},1},MM}
\end{bmatrix}
}^{-1}
({\bf y}_i-{\bf h}_{{\rm d},k})
.
\end{aligned}
\end{equation}
Note that one may adopt the LMMSE estimator to achieve better performance if the covariance of ${\bf h}_{{\rm u},k} ={\rm diag}({\bf h}_{{\rm r},1})^{-1}{\bf h}_{{\rm r},k}$ is available \cite[Section V]{Liuliang_CE2020TWC}.
In the next timeslot, the next $M$ IRS elements are switched on with $[\theta_{t,M+1},\theta_{t,M+2},\cdots,\theta_{t,2M}]^{\rm T}={\bf 1}$ while the other elements are switched off to estimate the next $M$ coefficients in ${\bf h}_{{\rm u},k}$. The estimation continues in this way until all the coefficients in ${\bf h}_{{\rm u},k} $ are estimated, which finally costs $I=\lceil \frac{N}{M}\rceil$ timeslots.
The overall pilot overhead of the protocol in \cite{Liuliang_CE2020TWC} is $K+N+\lceil \frac{N}{M}\rceil(K-1)$.

\subsection{Always-ON Channel Estimation Protocol}\label{proposed_protocol}
\begin{figure}
[!ht]
\centering
\includegraphics[width=.8\columnwidth]{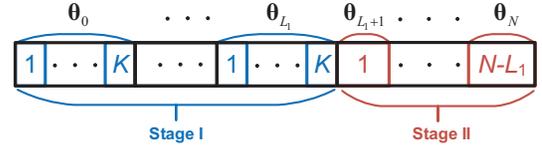}
\caption{Proposed always-ON protocol.}
\label{protocol:d}
\end{figure}

We propose a novel always-ON channel estimation protocol without switching off
selected IRS elements. The proposed protocol consists of two stages, as illustrated in {\figurename~\ref{protocol:d}}.
In particular, stage I contains $L_1+1$ timeslots, where $L_1 = \lceil \frac{N}{M}\rceil$, and each timeslot contains $K$ samples. Stage II contains $L_2=N-L_1$ timeslots, and each timeslot contains only one sample.
Therefore, we have $K(L_1+1)+L_2$ received samples in total. As defined in \eqref{equ:y_model1}, the $t$-th received sample is given by
\begin{equation}\label{equ:y_model2}
\begin{aligned}[b]
{\bf y}_{t}&=\sum_{k=1}^K \left({\bf h}_{{\rm d},k}+ {\bf G}^{\rm T} {\rm diag} ({\bm \theta}_t)  {\bf h}_{{\rm r},k} \right) x_{k,t}
+{ {\bf z}_t}
,
\end{aligned}
\end{equation}
where $t=1,2,\cdots,K(L_1+1)+L_2$.

\subsubsection{Received Signal Samples in Each Training Timeslot}
To facilitate analysis, we introduce new notations on the received signal samples within one training timeslot indexed by $\ell=0,1,\cdots,N$.
\begin{itemize}
\item
{{\bf {Stage I}} (Users send orthogonal pilot sequences ${\bf X}$)}:
Stage I consists of training timeslots $\ell=0,1,\cdots,L_1$.
Denote by ${\bf X} \in {\mathbb C}^{K \times K}$,  where ${\bf X}^{\rm H} {\bf X}=K{\bf I}_K$, the orthogonal pilot sequences consisting of unit-modulus elements.
In the $\ell$-th timeslot, $K$ users transmit ${\bf X}$ by $K$ samples, and the IRS is configured by the phase  shifting vector ${\bm \theta}_\ell$.
The $K$ received samples in the $\ell$-th timeslot are ${\bf y}_{K\ell+1},{\bf y}_{K\ell+2},\cdots,{\bf y}_{K\ell+K}$.
Let ${{\bf Y}}_{\ell}=[{\bf y}_{K\ell+1},{\bf y}_{K\ell+2},\cdots,{\bf y}_{K\ell+K}]$. We have
\begin{equation}\label{equ:y_model_stage_I}
\begin{aligned}[b]
{{\bf Y}}_{\ell}&=\left({\bf H}_{\rm d} + {\bf G}^{\rm T} {\rm diag} ({\bm \theta}_{\ell}) {\bf H}_{\rm r} \right)  {\bm X}
+{{\bf Z}}_{\ell}
,
\end{aligned}
\end{equation}
where
\begin{align}
{\bf H}_{\rm d}&=[{\bf h}_{{\rm d},1},\cdots,{\bf h}_{{\rm d},K}] \in {\mathbb C}^{M \times K},\\
{\bf H}_{\rm r}&=[{\bf h}_{{\rm r},1},\cdots,{\bf h}_{{\rm r},K}] \in {\mathbb C}^{N \times K}
\end{align}
are the stacked channel coefficient matrices,
$\ell$ is the timeslot index,
and ${{\bf Z}}_{\ell}=[{\bf z}_{K\ell+1},{\bf z}_{K\ell+2},\cdots,{\bf z}_{K\ell+K}]$ denotes the noise.

\item {{\bf {Stage II}} (Users send pilot $\bar{\bf x}$, which is the first column of $\bf X$)}:
Stage II consists of training timeslots $\ell=L_1+1,L_1+2,\cdots,N$.
We denote the first column of $\bf X$ by
\begin{equation}\label{equ:barx}
\bar{\bf x}=[{\bar x}_{1},\cdots,{\bar x}_{K}]^{\rm T}.
\end{equation}
In the $\ell$-th timeslot, the users  transmit  $\bar{\bf x}$, while the IRS is configured by ${\bm \theta}_\ell$.
The received signal in stage II is denoted by
\begin{equation}\label{equ:y_model_stage_III}
\begin{aligned}[b]
\bar{\bf y}_{\ell}
&= \left({\bf H}_{\rm d} + {\bf G}^{\rm T} {\rm diag} ({\bm \theta}_\ell) {\bf H}_{\rm r} \right)  \bar{\bf x}+{\bf z}_{\ell+(K-1)L_1+K}
,
\end{aligned}
\end{equation}
where $\bar{\bf y}_{\ell}={\bf y}_{\ell+(K-1)L_1+K}$, and $\ell=L_1+1,L_1+2,\cdots,N$.
\end{itemize}

$\bar{\bf y}_{\ell}$ may extend to all the $N+1$ timeslots in the protocol by
\begin{equation}\label{equ:bar_y}
\begin{aligned}[b]
\bar{\bf y}_{\ell}=
\begin{cases}
{\bf y}_{1+K \ell}, \; & {\rm for} \; \ell=0,\cdots,L_1,\\
{\bf y}_{\ell+(K-1)L_1+K}, & {\rm for} \; \ell=L_1+1,\cdots,N.
\end{cases}
\end{aligned}
\end{equation}

The pilot overhead in stage I is $(\lceil \frac{N}{M}\rceil+1) K$, and the overhead in stage II is $N-\lceil \frac{N}{M}\rceil $. The overall pilot overhead of the proposed protocol is $K+N+\lceil \frac{N}{M}\rceil(K-1)$.
Note that the pilot overhead is significantly reduced by about $M$ times compared to the protocol in
\cite{Araujo2021JSAC_CE_PARAFAC,Mishra2019CEonoff,Elbir2020WCL_DL_CE} and
\cite{Zhouzy2020decompositionCE,Kundu2021OJCSLMMSE_DFTGOOD,Alwazani2020OJCSLMMSE_DFT}, which require $NK$ pilots.

\subsubsection{Signal Pre-processing}
In the proposed protocol, we set ${\bm \theta}_0=-{\bm \theta}_1$ to decouple the estimation on the BS-user channel\footnote{
 The  BS-user channel can be estimated based on ${{\bf Y}}_0$ and ${{\bf Y}}_1$  by the linear minimum mean squared error  estimator, which is similar to the methods in existing works
\cite{Kundu2021OJCSLMMSE_DFTGOOD,Alwazani2020OJCSLMMSE_DFT,Liuliang_CE2020TWC} (See Appendix \ref{app_Estimate_Hd}).} and  on the cascaded IRS channel.
To facilitate the estimation of the cascaded IRS channel, signal pre-processing to remove the BS-user channel  from the received signals  is performed.

\begin{itemize}
\item  \emph{Pre-processing on ${{\bf Y}}_\ell$ for $\ell=1,2,\cdots,L_1$}:
For $\ell=1$, we have
\begin{equation}\label{equ:R1}
\begin{aligned}[b]
{\bf R}_{1} &= \frac{1}{2}\left({{\bf Y}}_1-{{\bf Y}}_0\right) \\
&={\bf G}^{\rm T} {\rm diag} ({\bm \theta}_1) {\bf H}_{\rm r}{\bf X}
+\tilde{\bf Z}_1
,
\end{aligned}
\end{equation}
where the elements in $\tilde{\bf Z}_1$ follow i.i.d. ${\cal{CN}}({ 0},\frac{1}{2}\sigma_0^2) $.
For $\ell=2,3,\cdots,L_1$, we have
\begin{equation}\label{equ:R2L1}
\begin{aligned}[b]
{\bf R}_{\ell} &= {{\bf Y}}_{\ell} -\frac{1}{2}\left({{\bf Y}}_0+{{\bf Y}}_1\right)\\
&={\bf G}^{\rm T} {\rm diag} ({\bm \theta}_\ell) {\bf H}_{\rm r} {\bf X}
+\tilde{\bf Z}_{\ell}
,
\end{aligned}
\end{equation}
where the elements in $\tilde{\bf Z}_{\ell}$ follow ${\cal{CN}}({ 0},\frac{3}{2}\sigma_0^2) $.
Note that the difference between \eqref{equ:R1} and \eqref{equ:R2L1} is that they have different noise variances.

\item \emph{Pre-processing on $\bar{\bf y}_{\ell}$ for $\ell=1,2,\cdots,N$}:
In the same manner, the BS-user channel is removed from $\bar{\bf y}_{\ell}$
for $\ell=1,2,\cdots,N$:
\begin{equation}\label{equ:rbar}
\begin{aligned}[b]
\bar{\bf r}_{\ell} &= \bar{\bf y}_{\ell}-\frac{1}{2}\left(\bar{\bf y}_{0}+\bar{\bf y}_{1}\right)\\
&={\bf G}^{\rm T} {\rm diag} ({\bm \theta}_\ell) {\bf H}_{\rm r}\bar{\bf x}
+\bar{\bf z}_{\ell}\\
&={\bf G}^{\rm T} {\rm diag} ({\bf H}_{\rm r}\bar{\bf x})  {\bm \theta}_\ell
+\bar{\bf z}_{\ell}
,
\end{aligned}
\end{equation}
where the elements in $\bar{\bf z}_{\ell}$ follow ${\cal{CN}}({ 0},\frac{3}{2}\sigma_0^2) $.
\end{itemize}

\begin{lemma}[Effectiveness of the proposed protocol]\label{lemma0}
The cascaded channels ${\bf H}_{{\rm I},k}={\bf G}^{\rm T} {\rm{diag}}({\bf h}_{{\rm r},k})$ ($k=1,2,\cdots,K$) for all $K$ users can be perfectly recovered with probability one by adopting an orthogonal phase shifting configuration matrix ${\bm \Phi}=[{\bm \theta}_1,{\bm \theta}_2,\cdots,{\bm \theta}_N]$ whose elements are all non-zero, if there is no noise, and
${\bf G}={\bf F}_{\rm R} {\ddot{\bf G}} {\bf F}_{\rm B}^{\rm T}$ and ${\bf h}_{{\rm r},k}={\bf F}_{\rm R}{\ddot{\bf h}}_{{\rm r},k}$ where ${\bf F}_{\rm B} \in {\mathbb C}^{M \times M}$ and ${\bf F}_{\rm R} \in {\mathbb C}^{N \times N}$ is respectively the angular domain basis for the BS antenna array and the IRS,
${\ddot{\bf G}}=[{\ddot{\bf g}}_1,{\ddot{\bf g}}_2,\cdots,{\ddot{\bf g}}_M]$, and ${\ddot{\bf g}}_m$ and ${\ddot{\bf h}}_{{\rm r},k}$ for all $m$ and $k$ are pairwise  independent following zero-mean multivariate normal distributions.
\end{lemma}

\begin{IEEEproof}
See Appendix \ref{proof_lemma0}.
\end{IEEEproof}

Compared with \cite{Liuliang_CE2020TWC} and \cite{double_IRS}, the proposed protocol has two main differences, which provide the opportunity to keep all the IRS elements ON and to reduce the pilot overhead by utilizing the common-link structure at the same time.
Here, we try to explain the intuition by supposing we adopt a similar estimation algorithm to those in \cite{Liuliang_CE2020TWC} and \cite{double_IRS}, i.e., first estimate the reference channel and then estimate the relative channels.
Firstly, instead of requiring a specific reference user, we design a virtual reference channel ${\bf H}_{\rm v}={\bf G}^{\rm T} {\rm diag} ({\bf H}_{\rm r}\bar{\bf x})$, which is fair for all users and can be reconstructed using the $N$ observations in \eqref{equ:rbar}.
Secondly, in our protocol, the relative channels can be estimated by using ${\bf R}_{\ell}$ in \eqref{equ:R1} and \eqref{equ:R2L1}
resulting a  measurement matrix
$[({\bf H}_{\rm v}{\rm diag} ({\bm \theta}_1))^{\rm T},\cdots,({\bf H}_{\rm v}{\rm diag} ({\bm \theta}_{L_1}))^{\rm T}]^{\rm T}$.
One can see that there are $L_1$ different ${\bm \theta}_{\ell}$ in the measurement matrix instead of a fixed one as in \cite{double_IRS}. When $L_1 \geq \lceil \frac{N}{M}\rceil$, it is possible that the rank of the  measurement matrix becomes $N$ with a proper phase shifting configuration to obtain a reasonable estimation.\footnote{
It is seen that the  selected on-off protocol in \cite{Liuliang_CE2020TWC} can be treated as a special case of our protocol by setting $\bar{\bf x}=[1,0,\cdots,0]^{\rm T}$ and selecting different parts of the IRS elements to be ON and OFF  to obtain $L_1$  different ${\bm \theta}_{\ell}$, as introduced in Section \ref{overview_onoff}.
}
Note that the above two-step channel estimation algorithm is only for explaining the intuition of the proposed approach. In the next section, we will propose an optimization-based cascaded channel estimation algorithm that may achieve more reliable  performance.

\section{Optimization-Based MU-Cascaded IRS Channel Estimation}\label{sec:opt_est}
In this section, we propose an optimization-based channel estimation on the cascaded IRS channel ${\bf h}_{{\rm I},k,m}$ for all $m$ and $k$ based on the pre-processed observations $\{{\bf R}_{\ell},\bar{\bf r}_{\ell}\}$ in \eqref{equ:R1}, \eqref{equ:R2L1} and \eqref{equ:rbar}.
We  consider a general decomposition on the cascaded channel, which is friendly in utilizing the channel prior knowledge
and the common-link structure across the multiple users.
Specifically, we adopt the MAP
approach to estimate
 the cascaded channel ${\bf h}_{{\rm I},k,m}$ for all $m$ and $k$
given the pre-processed observations $\{{\bf R}_{\ell},\bar{\bf r}_{\ell}\}$.
An alternating optimization algorithm with efficient initialization is further proposed to achieve a local optimum  of the MAP problem.


\subsection{MAP Problem Formulation}
As shown in \eqref{equ:cascaded_channel_vector}, the cascaded channel ${\bf h}_{{\rm I},k,m}$ can be decomposed by the common BS-IRS channel ${\bf g}_m$ and the IRS-user channel ${\bf h}_{{\rm r},k}$ as follows:
\begin{equation}\label{equ:cascaded_channel_vector_2v}
\begin{aligned}[b]
{\bf h}_{{\rm I},k,m}={\rm{diag}}({\bf h}_{{\rm r},k}) {\bf g}_m.
\end{aligned}
\end{equation}
However, the main challenge to estimate the individual ${\bf h}_{{\rm r},k}$ and ${\bf g}_m$  is that  in the MAP formulation, prior distributions of ${\bf g}_m$ and ${\bf h}_{{\rm r},k}$ will be needed, but
it is difficult to  obtain individual  covariances of ${\bf g}_m$ and ${\bf h}_{{\rm r},k}$ based on the covariance of the cascaded channel.\footnote{
One possible way to estimate the covariance of the cascaded channel ${\bf C}_m^{(k)}$ is using the
 the maximum likelihood estimator $\hat{\bf C}_m^{(k)}=\frac{1}{J} \sum_{j=1}^J \left[ \hat{\bf h}_{{\rm I},k,m}(j) \hat{\bf h}_{{\rm I},k,m}(j)^{\rm H} \right]$, where $\{\hat{\bf h}_{{\rm I},k,m}(j)\}$ are the estimated historical cascaded channels in the past $j=1,2,\cdots,J$ transmission frames.
  Note that similar covariances are also required by the LMMSE estimators for the selected on-off channel estimation protocol in \cite{Liuliang_CE2020TWC} (See equations (72) and (86) in \cite{Liuliang_CE2020TWC}).
}

To address this issue, we consider
a more general auxiliary variable set for the cascaded channel decomposition:
\begin{equation}\label{equ:set_opt}
\begin{aligned}[b]
{\cal A}=\left\{ \left.\left\{{\bf H}_{\rm g},{\bf H}_{\rm u}\right\} \right|
{\bf h}_{{\rm I},k,m}={\rm{diag}}({\bf h}_{{\rm u},k}) {\bf h}_{{\rm g},m}, \forall k, \forall m
\right\},
\end{aligned}
\end{equation}
where ${\bf H}_{\rm g}=[{\bf h}_{{\rm g},1},\cdots,{\bf h}_{{\rm g},M}]$ is the common-link variable whose $m$-th column is ${\bf h}_{{\rm g},m}$, and ${\bf H}_{\rm u}=[{\bf h}_{{\rm u},1},\cdots,{\bf h}_{{\rm u},K}]$ is the user-specific variable  whose $k$-th column is ${\bf h}_{{\rm u},k}$.
One may verify that $\{{\bf G}, {\bf H}_{\rm r}\}\in {\cal A}$ according to \eqref{equ:cascaded_channel_vector_2v}.
Based on this, we formulate an optimization problem on  ${\bf H}_{\rm g}$  and ${\bf H}_{\rm u}$ using the MAP approach, which is given by
\begin{equation*}
\begin{aligned}[b]
& 
{\mathcal{P}}{(\text{A})}
\quad \max_{ {\bf H}_{\rm g}, {\bf H}_{\rm u} } \;
f_{{\rm A}}({\bf H}_{\rm g}, {\bf H}_{\rm u})
,
\end{aligned}
\end{equation*}
where the objective function is given by\footnote{
One may substitute different prior setups to $p({\bf h}_{{\rm I},k,m} )$ in \eqref{equ:prior}.
In addition, if channel prior knowledge is unavailable, one may simply remove $p({\bf h}_{{\rm I},k,m} )$ from \eqref{equ:obj_f},
and the optimization becomes the maximum likelihood approach.
}
\begin{equation}\label{equ:obj_f}
\begin{aligned}[b]
&f_{{\rm A}}({\bf H}_{\rm g}, {\bf H}_{\rm u})
= \sum_{m=1}^M \sum_{k=1}^K \ln p({\bf h}_{{\rm I},k,m} ) \\
&\quad+\sum_{\ell=1}^{L_1} \ln p({\tilde{\bf R}}_{\ell}| {\bf H}_{\rm g}, {\bf H}_{\rm u} )
+\sum_{\ell=L_1+1}^{N} \ln p( \bar{\bf r}_{\ell} | {\bf H}_{\rm g}, {\bf H}_{\rm u} )
,
\end{aligned}
\end{equation}
and $\tilde{\bf R}_{\ell}= {{\bf R}}_{\ell} {\bf X}^{-1}$.
Note that $f_{{\rm A}}({\bf H}_{\rm g}, {\bf H}_{\rm u})$ in \eqref{equ:obj_f} only requires the prior distribution of ${\bf h}_{{\rm I},k,m}$, which is given by
\begin{equation}\label{equ:prior}
\begin{aligned}[b]
p({\bf h}_{{\rm I},k,m} )\propto e^{
- {\bf h}_{{\rm g},m}^{\rm H} {\rm{diag}}({\bf h}_{{\rm u},k})^{\rm H} {{\bf C}_m^{(k)}}^{-1} {\rm{diag}}({\bf h}_{{\rm u},k}) {\bf h}_{{\rm g},m}
},
\end{aligned}
\end{equation}
where $\propto$ denotes equality up to a
scaling that is independent of the variables (i.e., ${\bf h}_{{\rm I},k,m}$ for \eqref{equ:prior}).
The likelihood functions $p({\tilde{\bf R}}_{\ell}| {\bf H}_{\rm g}, {\bf H}_{\rm u})$ and $p( \bar{\bf r}_{\ell} | {\bf H}_{\rm g}, {\bf H}_{\rm u} )$ are given by
\begin{align}
p(\tilde{{\bf R}}_{\ell}| {\bf H}_{\rm g}, {\bf H}_{\rm u} )
&\propto
e^{-\sigma_{\ell}^{-2} \left\|
\tilde{{\bf R}}_{\ell}- {\bf H}_{\rm g}^{\rm T} {\rm diag} \left({\bm \theta}_{\ell}\right) {\bf H}_{{\rm u}}
\right\|^2_{\rm F} }
, \label{equ:likelihood_R}
\\
p( \bar{\bf r}_{\ell} | {\bf H}_{\rm g}, {\bf H}_{\rm u} )
&\propto
e^{-\bar\sigma_{\ell}^{-2} \left\|
\bar{\bf r}_{\ell}-  {\bf H}_{\rm g}^{\rm T} {\rm diag} \left({\bm \theta}_{\ell} \right) {\bf H}_{{\rm u}} \bar{\bf x}
\right\|^2_2 }
, \label{equ:likelihood_r}
\end{align}
where $\sigma_{1}^2=\frac{1}{2K}\sigma_0^2$, $\sigma_{\ell}^2=\frac{3}{2K}\sigma_0^2$ for $\ell=2,3,\cdots,L_1$,
and $\bar\sigma_{\ell}^2=\frac{3}{2}\sigma_0^2$ for $\ell=L_1+1,L_1+2,\cdots,N$.
Finally, after dropping all the irrelevant constant terms, the objective function is equivalently written as
\begin{equation}\label{equ:obj_f_eq}
\begin{aligned}[b]
&f_{{\rm A}}({\bf H}_{\rm g}, {\bf H}_{\rm u})
=-\sum_{\ell=1}^{L_1}
{\frac{1}{\sigma_{\ell}^{2}} \left\|
\tilde{{\bf R}}_{\ell}- {\bf H}_{\rm g}^{\rm T} {\rm diag} \left({\bm \theta}_{\ell}\right) {\bf H}_{{\rm u}}
\right\|^2_{\rm F} }\\
&\quad-\sum_{\ell=L_1+1}^{N}
{ \frac{1}{\bar\sigma_{\ell}^{2}} \left\|
\bar{\bf r}_{\ell}-{\bf H}_{\rm g}^{\rm T} {\rm diag} \left({\bm \theta}_{\ell} \right) {\bf H}_{{\rm u}} \bar{\bf x}
\right\|^2_2 }
\\
&\quad-\sum_{m=1}^M \sum_{k=1}^K {\bf h}_{{\rm g},m}^{\rm H} {\rm{diag}}({\bf h}_{{\rm u},k})^{\rm H} {{\bf C}_m^{(k)}}^{-1} {\rm{diag}}({\bf h}_{{\rm u},k}) {\bf h}_{{\rm g},m}  .
\end{aligned}
\end{equation}
Note that ${\mathcal{P}}{(\text{A})}$ does not have a unique solution but all the solutions are equivalent for the purpose of estimation of the cascaded channel ${\bf h}_{{\rm I},k,m}$ for all $m$ and $k$.
\begin{lemma}[Equivalence of the solution of ${\mathcal{P}}{(\text{A})}$]\label{eq_decomposite}
Let ${\bf H}_{\rm g}^\star$ and ${\bf H}_{\rm u}^\star$ be an optimal solution of ${\mathcal{P}}{(\text{A})}$, then ${\rm diag} \left({\bf a}\right){\bf H}_{\rm g}^\star$ and ${\rm diag} \left({\bf a}\right)^{-1}{\bf H}_{\rm u}^\star$ is also an optimal solution of ${\mathcal{P}}{(\text{A})}$
for any coefficient ${\bf a}=[a_1,a_2,\cdots,a_N]^{\rm T}$ with $|a_n|\neq0$ for all $n=1,2,\cdots,N$.
\end{lemma}

\begin{IEEEproof}
See Appendix \ref{proof_lemma1}.
\end{IEEEproof}

As a result, there is ambiguity in estimating individual channels from solving ${\mathcal{P}}{(\text{A})}$. Nevertheless, the cascaded channel is unique regardless of the coefficient ${\bf a}$.

%


\subsection{ Channel Estimation Algorithm based on Alternative Optimization}\label{sec:opt_est_AO}

Solving ${\mathcal{P}}{(\text{A})}$ is difficult  due to the optimization variables being coupled in the likelihood functions \eqref{equ:likelihood_R} and \eqref{equ:likelihood_r}.
Fortunately, we will show that ${\mathcal{P}}{(\text{A})}$ is actually bi-convex (see Lemma \ref{fu_convex} and Lemma \ref{fg_convex}), which can be solved by alternative optimization. In particular, we decompose ${\mathcal{P}}{(\text{A})}$ into two convex sub-problems, and
the optimal solutions for these sub-problems will be derived accordingly.

\subsubsection{Optimize ${\bf H}_{\rm u}$}
We investigate the optimization of ${\bf H}_{\rm u}$ while ${\bf H}_{\rm g}$ are fixed.
After dropping all irrelevant terms, the sub-problem  is given by
\begin{align*}
{\mathcal{P}}{({\text A}_{{\rm u}})} \quad \min_{ {\bf H}_{\rm u} }\; f_{{\rm u}}({\bf H}_{\rm u})
,
\end{align*}
where
\begin{equation}\label{equ:obj_f_hrk}
\begin{aligned}[b]
&f_{{\rm u}}({\bf H}_{\rm u})
=\sum_{\ell=1}^{L_1}
{\frac{1}{\sigma_{\ell}^{2}} \left\|
\tilde{{\bf R}}_{\ell}- {\bf D}_{\ell} {\bf H}_{{\rm u}}
\right\|^2_{\rm F} }\\
&\quad+\sum_{\ell=L_1+1}^{N}
{ \frac{1}{\bar\sigma_{\ell}^{2}} \left\|
\bar{\bf r}_{\ell}-{\bf D}_{\ell} {\bf H}_{{\rm u}} \bar{\bf x}
\right\|^2_2 }
+\sum_{k=1}^K {\bf h}_{{\rm u},k}^{\rm H} {\bf C}_{{\rm u},k} {\bf h}_{{\rm u},k}
,
\end{aligned}
\end{equation}
and
\begin{align}
{\bf D}_{\ell}&={\bf H}_{\rm g}^{\rm T} {\rm diag} \left({\bm \theta}_{\ell}\right), \; {\text{for}} \; \ell=1,2,\cdots,N,\\
{\bf C}_{{\rm u},k}&=\sum_{m=1}^M {\rm{diag}}({\bf h}_{{\rm g},m})^{\rm H} {{\bf C}_m^{(k)}}^{-1} {\rm{diag}}({\bf h}_{{\rm g},m}), \quad \forall  k.
\end{align}

\begin{lemma}[Convexity of ${\mathcal{P}}{({\text A}_{{\rm u}})}$]\label{fu_convex}
For any fixed ${\bf H}_{\rm g}$, the objective function of ${\mathcal{P}}{({\text A}_{{\rm u}})}$ is a convex quadratic function of the vectorization of ${\bf H}_{\rm u}$, which is denoted by ${\rm{vec}}({\bf H}_{\rm u})$.
\end{lemma}

\begin{IEEEproof}
See Appendix \ref{proof_lemmafu}.
\end{IEEEproof}

Based on Lemma \ref{fu_convex}, the optimal solution of ${\mathcal{P}}{({\text A}_{{\rm u}})}$ is the root of the first order derivative of $f_{{\rm u}}({\rm{vec}}({\bf H}_{\rm u}))$,  which is given by
\begin{equation}\label{equ:opt_hrk}
\begin{aligned}[b]
{\rm{vec}}({\bf H}_{\rm u}^\star)
&= {\bm \Lambda}_{{\rm u}}^{-1} {\bm \nu}_{{\rm u}}
,
\end{aligned}
\end{equation}
where
\begin{align}
{\bm \Lambda}_{{\rm u}}&=
 {\bf C}_{{\rm u}}
+{\bf I}_K \otimes \left(\sum_{\ell=1}^{L_1} \frac{1}{\sigma_{\ell}^{2}} {\bf D}_{\ell}^{\rm H} {\bf D}_{\ell}\right) \notag\\
&\quad + \left(\bar{\bf x}^\ast \bar{\bf x}^{\rm T}\right)\otimes \left(\sum_{\ell=L_1+1}^{N} \frac{1}{\bar\sigma_{\ell}^{2}} |\bar{x}_k|^2 {\bf D}_{\ell}^{\rm H} {\bf D}_{\ell}\right)
, \label{equ:opt_hrk_e1}\\
{\bf C}_{{\rm u}}&={\rm{blkdiag}}({\bf C}_{{\rm u},1},{\bf C}_{{\rm u},2},\cdots,{\bf C}_{{\rm u},K}),
\end{align}
and
\begin{equation}\label{equ:opt_hrk_e2}
\begin{aligned}[b]
{\bm \nu}_{{\rm u}}
&={\rm{vec}}\left(
\sum_{\ell=1}^{L_1} \frac{1}{\sigma_{\ell}^{2}} {\bf D}_{\ell}^{\rm H}
\tilde{{\bf R}}_{\ell}
+ \sum_{\ell=L_1+1}^{N} \frac{1}{\bar\sigma_{\ell}^{2}}
 {\bf D}_{\ell}^{\rm H} \bar{\bf r}_{\ell} \bar{\bf x}^{\rm H}
\right).
\end{aligned}
\end{equation}

\subsubsection{Optimize ${\bf H}_{{\rm g}}$}
Similarly, the sub-problem of optimizing ${\bf H}_{{\rm g}}$ is given by
\begin{align*}
{\mathcal{P}}{(\text{A}_{{\rm g}})} \; \min_{ {\bf H}_{{\rm g}} }\; \sum_{m=1}^M f_{{\rm g},m}({\bf h}_{{\rm g},m})
,
\end{align*}
where
\begin{align}
&f_{{\rm g},m}({\bf h}_{{\rm g},m})
=
 \sum_{k=1}^K \sum_{\ell=1}^{L_1} \frac{1}{\sigma_{\ell}^{2}} \left\|
\tilde{{r}}_{\ell,m,k}-{\bf h}_{{\rm g},m}^{\rm T}  {\bf b}_{\ell,k}
\right\|^2 \notag\\
&+ \sum_{\ell=L_1+1}^{N}\frac{1}{\bar\sigma_{\ell}^{2}}   \left\|
\bar{r}_{\ell,m}-\sum_{k=1}^K \bar{x}_k {\bf h}_{{\rm g},m}^{\rm T} {\bf b}_{\ell,k}
\right\|^2+{\bf h}_{{\rm g},m}^{\rm H} {\bf C}_{{\rm g},m} {\bf h}_{{\rm g},m}
, \label{equ:obj_f_gm}\\
&\quad{\bf b}_{\ell,k}={\rm diag}\left({\bm \theta}_{\ell}\right) {\bf h}_{{\rm u},k}, \; {\text{for}} \; \ell=1,2,\cdots,N,\\
&\quad{\bf C}_{{\rm g},m}=\sum_{k=1}^K {\rm{diag}}({\bf h}_{{\rm u},k})^{\rm H} {{\bf C}_m^{(k)}}^{-1} {\rm{diag}}({\bf h}_{{\rm u},k}), \quad \forall  m,
\end{align}
$\tilde{{r}}_{\ell,m,k}$ denotes the entry in the $m$-th row and $k$-th column of $\tilde{\bf R}_{\ell}$,
and $\bar{r}_{\ell,m}$ denotes the $m$-th entry in $\bar{\bf r}_{\ell}$.

\begin{lemma}[Convexity of ${\mathcal{P}}{({\text A}_{{\rm g}})}$]\label{fg_convex}
For any fixed ${\bf H}_{\rm u}$, the objective function of ${\mathcal{P}}{({\text A}_{{\rm g}})}$ is a convex quadratic function of $\{{\bf h}_{{\rm g},1},{\bf h}_{{\rm g},2},\cdots,{\bf h}_{{\rm g},M}\}$.
\end{lemma}

\begin{IEEEproof}
See Appendix \ref{proof_lemmafg}.
\end{IEEEproof}

Based on lemma \ref{fg_convex}, the optimal ${\bf h}_{{\rm g},m}$ is the
root of the first order derivative of $f_{{\rm g},m}({\bf h}_{{\rm g},m})$, which is given by
\begin{equation}\label{equ:opt_gm}
\begin{aligned}[b]
{\bf h}_{{\rm g},m}^\star
&= {\bm \Lambda}_{{\rm g},m}^{-1} {\bm \nu}_{{\rm g},m}
,
\end{aligned}
\end{equation}
for $m=1,2,\cdots,M$, where
\begin{equation}\label{equ:opt_gm_e1}
\begin{aligned}[b]
{\bm \Lambda}_{{\rm g},m}&={\bf C}_{{\rm g},m}^{-1}
+\sum_{k=1}^K\sum_{\ell=1}^{L_1} \frac{1}{\sigma_{\ell}^{2}} {\bf b}_{\ell,k}^\ast {\bf b}_{\ell,k}^{\rm T}\\
&\;+ \sum_{\ell=L_1+1}^{N}\frac{1}{\bar\sigma_{\ell}^{2}}
\left(\sum_{k =1}^K \bar{x}_k {\bf b}_{\ell,k}^{\rm T}\right)^{\rm H}
\left(\sum_{k =1}^K \bar{x}_k {\bf b}_{\ell,k}^{\rm T}\right)
,
\end{aligned}
\end{equation}
and
\begin{equation}\label{equ:opt_gm_e2}
\begin{aligned}[b]
&{\bm \nu}_{{\rm g},m}=\sum_{k=1}^K\sum_{{\ell}=1}^{L_1} \frac{\tilde{{r}}_{\ell,m,k}}{\sigma_{\ell}^{2}} {\bf b}_{\ell,k}^\ast
+ \sum_{\ell=L_1+1}^{N}\frac{\bar{r}_{\ell,m}}{\bar\sigma_{\ell}^{2}}
\left(\sum_{k =1}^K \bar{x}_k^{\ast} {\bf b}_{\ell,k}^{\ast}\right)
.
\end{aligned}
\end{equation}

\subsubsection{Initial Estimation on ${\bf H}_{{\rm g}}$}
 The quality of the solution obtained by the alternative optimization  depends heavily on the initial point.
 Here, we  propose
an efficient estimator for ${\bf H}_{{\rm g}}$  to initialize the proposed alternative optimization algorithm.
In particular, we construct a special  $\left\{{\bf H}_{\rm g},{\bf H}_{\rm u}\right\}$ pair whose elements are given by
\begin{align}
{\bf h}_{{\rm g},m}&= {\rm diag} ({\bf H}_{\rm r}\bar{\bf x}) {\bf g}_m, \label{equ:hg1}\\
{\bf h}_{{\rm u},k}&={\rm diag} ({\bf H}_{\rm r}\bar{\bf x})^{-1}{\bf h}_{{\rm r},k}. \label{equ:hu1}
\end{align}
Substituting the above ${\bf H}_{\rm g}$  into \eqref{equ:rbar},  we have
\begin{equation}\label{equ:rbar_v2}
\begin{aligned}[b]
\bar{\bf r}_{\ell}
&={\bf H}_{\rm g}^{\rm T}  {\bm \theta}_\ell
+\bar{\bf z}_{\ell}
,
\end{aligned}
\end{equation}
for $\ell=1,2,\cdots,N$.
Therefore, ${\bf H}_{\rm g}$ can be initialized by the least squares (LS) estimator as follows:
\begin{equation}\label{equ:est_Hc}
\begin{aligned}[b]
\hat{\bf H}_{\rm g}=\left( \left[\bar{\bf r}_1,\cdots,\bar{\bf r}_N \right]
\left[{\bm \theta}_1,\cdots,{\bm \theta}_N \right]^{-1} \right)^{\rm T}.
\end{aligned}
\end{equation}
Since $\hat{\bf H}_{\rm g}$ in \eqref{equ:est_Hc} is unbiased and it has exploited most of the available observations in all the $N+1$ training timeslots,
it will give a good initial point.

\subsection{The Overall Proposed Algorithm}
The overall proposed cascaded IRS channel estimation algorithm  is summarized in Algorithm \ref{alg:P1}.
The convergence of the proposed alternating optimization algorithm is analyzed in Lemma \ref{convergence}.

\begin{algorithm}[!ht]
\caption{ Proposed alternating optimization algorithm for the  cascaded IRS channel estimation}
\label{alg:P1}
\begin{algorithmic}[1]
\STATE {Initialize ${\bf H}_{{\rm g}}$  by \eqref{equ:est_Hc}.}\\
\REPEAT
\STATE  Update ${\bf H}_{{\rm u}}$ by \eqref{equ:opt_hrk};
\STATE  Update ${\bf h}_{{\rm g},m}$ by \eqref{equ:opt_gm}  for all $m$;
\UNTIL{ $f_{{\rm A}}({\bf H}_{\rm g}, {\bf H}_{\rm u})$ in \eqref{equ:obj_f_eq} converges;}\\
\STATE {Output the cascaded channel ${\hat{\bf h}}_{{\rm I},k,m}={\rm{diag}}({\bf h}_{{\rm u},k}) {\bf h}_{{\rm g},m}$ for all $k$ and $m$.}
\end{algorithmic}
\end{algorithm}

\begin{lemma}[Convergence of the Proposed  Alternating Optimization Algorithm]\label{convergence}
The objective function $f_{{\rm A}}({\bf H}_{\rm g}, {\bf H}_{\rm u})$ is non-increasing in every step when ${\bf H}_{{\rm u}}$ or ${\bf H}_{{\rm g}}$ are updated, and the optimization  iterations in \eqref{equ:opt_hrk} and \eqref{equ:opt_gm}  converge to a local optimum
of ${\mathcal{P}}{(\text{A})}$.
\end{lemma}

\begin{IEEEproof}
As shown in Section \ref{sec:opt_est_AO}, the original problem is decomposed into two unconstrained minimization problems whose objectives are convex quadratic functions, and each subproblem has a unique optimal solution, which is derived in \eqref{equ:opt_hrk} and \eqref{equ:opt_gm}.
Therefore, the  whole alternating optimization algorithm will converge to  a local optimum of the original problem ${\mathcal{P}}{(\text{A})}$ \cite{BCD}.
\end{IEEEproof}

\emph{Remark}:
The complexity for updating ${\bf H}_{{\rm u}}$ by \eqref{equ:opt_hrk} is ${\cal O}(K^3 N^3+KMN^2)$.  The complexity for updating ${\bf h}_{{\rm g},m}$ by  \eqref{equ:opt_gm} is ${\cal O}(KN^3)$, and thus the complexity to update  ${\bf H}_{{\rm g}}$ is ${\cal O}(KMN^3)$. Therefore, the overall complexity of the solution is ${\cal O}(IK^3 N^3+IKMN^3)$, where $I$ denotes the number of iterations of the alternating optimization algorithm.\footnote{We will show in simulation that the algorithm will converge quickly in about two or three iterations. In addition, the ${\cal O}(N^3)$ complexity is costed by the matrix inversion operation. However, since the matrices required inversion operation are all Hermitian positive semi-definite matrices, they can be implemented very efficiently by advanced algorithms such as the Cholesky-decomposition-based algorithm \cite{matrix_inverse}.  }

\section{Training Phase Shifting Configuration}\label{sec:Phaseshift_cofig}
\subsection{ Motivation of the Phase Shifting Configuration}
\begin{figure}
[!ht]
\centering
\includegraphics[width=.9\columnwidth]{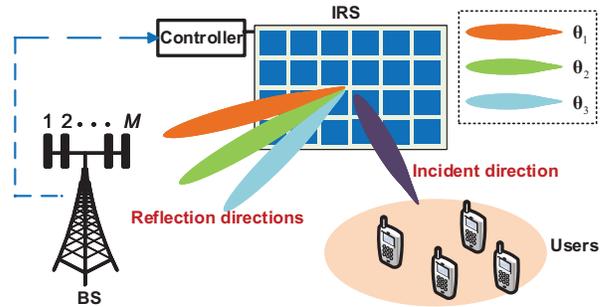}
\caption{Illustration of the impact of different phase shiftings.}
\label{theta_opt}
\end{figure}

The IRS steers the incident signal to different directions by configuring different phase shifting vectors ${\bm \theta}_{\ell}$, as illustrated in {\figurename~\ref{theta_opt}}.
According to the protocol, the cascaded channel is scanned by $N$ spatial directions in $N$ training timeslots, and a proper design on $\{ {\bm \theta}_1,{\bm \theta}_2,\cdots,{\bm \theta}_N \}$ guarantees that the whole channel information in all directions is contained by the received signals such that good channel estimation performance can be achieved.

In the SU-MISO scenario, the overall received measurements after removing the pilots and the BS-user channels is given by
\begin{equation}\label{equ:y_su}
\begin{aligned}[b]
{\bf Y}_{\rm{SU}}&= {\bf H}_{{\rm I},1} {\bm \Phi}
+{\bf Z}
,
\end{aligned}
\end{equation}
where ${\bm \Phi}=[{\bm \theta}_1,{\bm \theta}_2,\cdots,{\bm \theta}_N]$. The LS estimator may adopted as follows \cite{Zhouzy2020decompositionCE}:
\begin{equation}\label{equ:hatH_su}
\begin{aligned}[b]
\hat{\bf H}_{{\rm I},1}&= {\bf Y}_{\rm{SU}} {\bm \Phi}^{-1}
,
\end{aligned}
\end{equation}
where $\hat{\bf H}_{{\rm I},1}$ denotes the estimated cascaded channel.
Then ${\bm \Phi}$ is optimized by minimizing the MSE:
\begin{align*}
\min_{ {\bm \Phi} }\; & {\rm{tr}} \left(\left({\bm \Phi} {\bm \Phi}^{\rm H}\right)^{-1}\right)\\
{\bf s.t.} \;
& |{\bm \Phi}_{i,j}|=1, \quad \forall i,j=1,2,\cdots,N
.
\end{align*}
It is proved in \cite{Zhouzy2020decompositionCE} that the optimal value of the MSE is $1$, which can be achieved by the DFT matrix such that ${\bm \Phi}={\bf F}$, where
\begin{equation}\label{equ:DFT_F}
\begin{aligned}[b]
{\bf F}
=
\begin{bmatrix}
    1      &   1 &  1  & \cdots & 1 \\
    1 & e^{-\jmath 2 \pi \frac{1}{N}} & e^{-\jmath 2 \pi \frac{2}{N}} & \cdots & e^{-\jmath 2 \pi \frac{N-1}{N}} \\
    1 & e^{-\jmath 2 \pi \frac{2}{N}} & e^{-\jmath 2 \pi \frac{4}{N}} & \cdots & e^{-\jmath 2 \pi \frac{2(N-1)}{N}} \\
    \vdots     &  \vdots &  \vdots  & \ddots & \vdots \\
   1 & e^{-\jmath 2 \pi \frac{N-1}{N}} & e^{-\jmath 2 \pi \frac{2(N-1)}{N}} & \cdots & e^{-\jmath 2 \pi \frac{(N-1)^2}{N}}
\end{bmatrix}
.
\end{aligned}
\end{equation}

For the protocol extended from the SU case \cite{Zhouzy2020decompositionCE,Kundu2021OJCSLMMSE_DFTGOOD,Alwazani2020OJCSLMMSE_DFT} shown in {\figurename~\ref{protocol_tran}}, the transmit signals from users are the same in different timeslots.
Therefore, the columns of ${\bf F}$ may be permuted to any orders, and the MSE will remain the same.
However, in our proposed protocol in {\figurename~\ref{protocol:d}}, the transmit signals are different in stage I and stage II.
In particular, the received signals in stage I contribute to the estimation on both the common-link variable and the user-specific variables in the cascaded channels, while the signals in stage II contribute to the common-link variable only.
Hence, the  phase shifting vectors $ {\bm \theta}_1, {\bm \theta}_2,\cdots,{\bm \theta}_{L_1} $
in stage I require additional design.





\subsection{Optimization Formulation of the Phase Shifting Configuration for MU-MISO IRS Systems}

As shown in \eqref{equ:est_Hc}, the initial estimation on the common-link variable $\hat{\bf H}_{\rm g}$ in \eqref{equ:est_Hc}  is almost the same as the estimator for the SU case shown in \eqref{equ:hatH_su}.
Therefore, we still adopt the DFT-based phase shifting configuration for all the $N$ timeslots. Additionally, an additional steering direction ${\bm{\vartheta}} \in {\mathbb C}^{N \times 1}$ is introduced for a more flexible design:
\begin{equation}\label{equ:prop_phi}
\begin{aligned}[b]
{\bm \Phi}&= {\rm{diag}}({\bm{\vartheta}}){\bf F}
.
\end{aligned}
\end{equation}
Denote by $\vartheta_n$ the $n$-th element in ${\bm{\vartheta}}$. We have $|\vartheta_n|=1$ for all $n=1,2,\cdots,N$.
One can see that the value of ${\rm{tr}} \left(\left({\bm \Phi} {\bm \Phi}^{\rm H}\right)^{-1}\right)$ is kept at $1$ for any ${\bm{\vartheta}}$.
Denote by ${\bf f}_{\ell}$ the $\ell$-th column of $\bf F$. The training phase shifting vector in the $\ell$-th timeslot is given by
\begin{equation}\label{equ:prop_theta_ell}
\begin{aligned}[b]
{\bm \theta}_\ell&= {\rm{diag}}({\bm{\vartheta}}) {\bf f}_{\ell}
,
\end{aligned}
\end{equation}
where $\ell=1,2,\cdots,N$.
The remaining task is to design ${\bm{\vartheta}}$.

However, it is difficult to design a straightforward objective function to optimize ${\bm{\vartheta}}$ since the MSE of the estimated ${\bf H}_{\rm u}$ by the proposed algorithm is complicated.
Considering that we have the knowledge of the covariances ${\bf C}_m^{(k)}$ of the cascaded channels for all $m$ and $k$,
the  average received power of the effective IRS channel from user $k$ to the $m$-th BS antenna in timeslot $\ell$  can be denoted by a function of ${\bm{\vartheta}}$, ${\bf f}_{\ell}$ and ${\bf C}_m^{(k)}$:
\begin{equation}\label{equ:channel_gain}
\begin{aligned}[b]
Q_{\ell,k,m} &= {\mathbb E} \left[ | {\bf g}_m^{\rm T} {\rm{diag}}({\bm \theta}_\ell) {\bf h}_{{\rm r},k} |^2  \right]\\
 &= {\mathbb E} \left[ | {\bf h}_{{\rm I},k,m}^{\rm T} {\bm \theta}_\ell |^2  \right]\\
&= {\mathbb E} \left[   {\bm \theta}_\ell^{\rm H}
\left( {\bf h}_{{\rm I},k,m} {\bf h}_{{\rm I},k,m}^{\rm H} \right)^{\ast}
{\bm \theta}_\ell  \right]\\
&= {\bm \theta}_\ell^{\rm H}
\left( {\bf C}_m^{(k)} \right)^{\ast}
{\bm \theta}_\ell\\
&={\bm{\vartheta}}^{\rm H} {\rm{diag}}({\bf f}_{\ell})^{\rm H}
\left( {\bf C}_m^{(k)} \right)^{\ast}
{\rm{diag}}({\bf f}_{\ell}){\bm{\vartheta}}
.
\end{aligned}
\end{equation}
Since ${\bf f}_{\ell}$ and ${\bf C}_m^{(k)}$ are known variables, the summation of $Q_{\ell,k,m}$ over antennas $m=1,2,\cdots,M$, users $k=1,2,\cdots,K$ and timeslots $\ell=1,2,\cdots,L_1$ is a function of ${\bm{\vartheta}}$, which is given by
\begin{equation}\label{equ:channel_gain_all}
\begin{aligned}[b]
f_{\rm B}({\bm{\vartheta}})&=\sum_{\ell=1}^{L_1} \sum_{k=1}^K \sum_{m=1}^M Q_{\ell,k,m} \\
&=\sum_{\ell=1}^{L_1} \sum_{k=1}^K \sum_{m=1}^M
{\bm{\vartheta}}^{\rm H} {\rm{diag}}({\bf f}_{\ell})^{\rm H}
\left( {\bf C}_m^{(k)} \right)^{\ast}
{\rm{diag}}({\bf f}_{\ell}){\bm{\vartheta}} \\
&={\bm{\vartheta}}^{\rm H} \left(\sum_{\ell=1}^{L_1} \sum_{k=1}^K \sum_{m=1}^M
 {\rm{diag}}({\bf f}_{\ell})^{\rm H}
\left( {\bf C}_m^{(k)} \right)^{\ast}
{\rm{diag}}({\bf f}_{\ell})\right){\bm{\vartheta}} .
\end{aligned}
\end{equation}
We define
\begin{equation}
\begin{aligned}[b]
{\bf E}=\left(\sum_{\ell=1}^{L_1} \sum_{k=1}^K \sum_{m=1}^M
 {\rm{diag}}({\bf f}_{\ell})^{\rm H}
\left( {\bf C}_m^{(k)} \right)^{\ast}
{\rm{diag}}({\bf f}_{\ell})\right).
\end{aligned}
\end{equation}
The optimization problem on ${\bm{\vartheta}}$ is formulated to maximize $f_{\rm B}({\bm{\vartheta}})$:
\begin{equation*}
\begin{aligned}[b]
{\mathcal{P}}{(\text{B})}\quad \max_{ {\bm{\vartheta}} } \; &
f_{\rm B}({\bm{\vartheta}})={\bm{\vartheta}}^{\rm H} {\bf E} {\bm{\vartheta}}\\
{\bf s.t.} \;
& |\vartheta_n|=1, \quad \forall n=1,2,\cdots,N.
\end{aligned}
\end{equation*}


\subsection{Solution for ${\mathcal{P}}{(\text{B})}$}
${\mathcal{P}}{(\text{B})}$ is a non-convex problem due to the maximizing of a convex objective function and the unit-modulus constraints.
We solve ${\mathcal{P}}{(\text{B})}$ by the successive convex approximation (SCA) algorithm.
In particular, a surrogate problem, shown as follows, is iteratively solved:
\begin{equation*}
\begin{aligned}[b]
{\mathcal{P}}{({\text{B}}_i)}\quad \max_{ {\bm{\vartheta}} } \; &
{f}_{\rm B}^{(i)} ({\bm{\vartheta}},\bar{\bm{\vartheta}})\\
{\bf s.t.} \;
& |\vartheta_n|=1, \quad \forall n=1,2,\cdots,N,
\end{aligned}
\end{equation*}
where $i$ is the iteration index, $\bar{\bm{\vartheta}}$ is the solution of the surrogate problem in the $(i-1)$-th iteration, and ${f}_{\rm B}^{(i)}({\bm{\vartheta}},\bar{\bm{\vartheta}})$ is the first-order approximation of  ${f}_{\rm B}({\bm{\vartheta}})$ at $\bar{\bm{\vartheta}}$:
\begin{equation}\label{equ:surrogate}
\begin{aligned}[b]
{f}_{\rm B}^{(i)} ({\bm{\vartheta}},\bar{\bm{\vartheta}})
&= 2 {\rm Re} \left\{{\bar{\bm \vartheta}}^{\rm H} {\bf E} {\bm \vartheta}\right\}-{\bar{\bm \vartheta}}^{\rm H} {\bf E} {\bar{\bm \vartheta}}.
\end{aligned}
\end{equation}
One can see that ${f}_{\rm B}^{(i)}({\bm{\vartheta}},\bar{\bm{\vartheta}})$ is a linear function of ${\bm{\vartheta}}$, and thus the optimal solution of ${\mathcal{P}}{({\text{B}}_i)}$ is given by
\begin{equation}\label{equ:opt_theta}
\begin{aligned}[b]
{\bm{\vartheta}}=e^{\jmath \angle ({\bf E}{\bar{\bm \vartheta}})}.
\end{aligned}
\end{equation}
The proof on the convergence of the SCA algorithm can be referred to in \cite{SCA}.

\section{Numerical Examples}\label{simulation}
\subsection{Simulation Setups}
\begin{figure}
[!ht]
\centering
\includegraphics[width=.8\columnwidth]{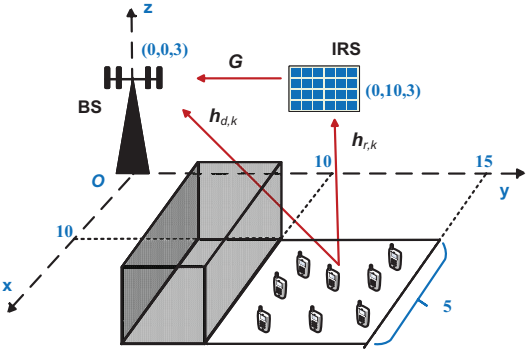}
\caption{The simulated IRS-aided $K$-user MISO communication scenario comprising of one $M$-antenna BS and one $N$-element IRS.}
\label{indoor_8user}
\end{figure}

This section evaluates the performance of the proposed cascaded channel estimation algorithm.
In particular, we consider the indoor femtocell network illustrated in {\figurename~\ref{indoor_8user}} in which $K=8$ users are randomly distributed in a 5 m $\times$ 5 m square area and are served by one BS and one IRS.
We generate the channel coefficients according to the 3GPP ray-tracing model \cite[Section 7.5]{3GPP} using the model parameters for the Indoor-Office scenario \cite[Table 7.5-6]{3GPP}.
The system parameters for the simulations are summarized in Table \ref{table_sim}, in which the path-loss is set according to the Indoor-Office pathloss model in \cite[Table 7.4.1-1]{3GPP}.

\begin{table}[!ht]
\footnotesize
\renewcommand{\arraystretch}{1.3}
\caption{Simulation Parameters}
\label{tablepm}
\centering
\begin{tabular}{c|c}
\hline
Parameters & Values \\
\hline
Carrier frequency & 2.4 GHz\\
\hline
 Transmission bandwidth  & $200$ kHz\\
\hline
Noise power spectral density & $-170$ dBm/Hz\\
\hline
Path-loss for BS-IRS and IRS-user links (dB)& $40 + 17.3 \lg d$\\
\hline
Path-loss for BS-user link (dB)& $30 + 31.9 \lg d+\zeta$\\
\hline
Penetration loss $\zeta$ due to obstacle  & 20 dB \\
\hline
 Reflection efficiency  of IRS & 0.8\\
\hline
Height of users & 1.5 m\\
\hline
Location of BS & (0, 0, 3m)\\
\hline
Location of IRS & (0, 10m, 3m)\\
\hline
\end{tabular}
\label{table_sim}
\end{table}

In the simulation, we consider two baseline schemes to benchmark the proposed scheme.
\begin{itemize}
\item {\bf Baseline 1 [LMMSE using the protocol in {\figurename~\ref{protocol:b}} \cite{Kundu2021OJCSLMMSE_DFTGOOD,Alwazani2020OJCSLMMSE_DFT}]}: This curve illustrates the performance of the LMMSE estimator proposed in \cite{Kundu2021OJCSLMMSE_DFTGOOD} and \cite{Alwazani2020OJCSLMMSE_DFT}. For simplicity, we assume that the BS-user channels have already been perfectly estimated by this scheme.
    The protocol illustrated in {\figurename~\ref{protocol:b}} is adopted.
    In additon, for fair comparison, the number of training timeslots is set as $\lceil \frac{N-1}{K}\rceil+\lceil \frac{N}{M}\rceil$ such that the total pilot overhead is just slightly higher than that of the proposed scheme.
\item {\bf Baseline 2 [Bilinear alternating least squares (BALS) algorithm \cite{Araujo2021JSAC_CE_PARAFAC}]}:
 In \cite{Araujo2021JSAC_CE_PARAFAC}, an iterative algorithm is proposed to estimate ${\bf G}$ and ${\bf H}_{\rm r}$ by utilizing the PARAFAC decomposition, which adopts the same channel estimation protocol as Baseline 1. Note that due to the ambiguity issue (see Lemma \ref{eq_decomposite} or \cite[Section IV]{Araujo2021JSAC_CE_PARAFAC}), the BALS also cannot exactly reconstruct ${\bf G}$ and ${\bf H}_{\rm r}$, and the actually estimated variable is still the cascaded channel.
\item {\bf Baseline 3 [MAP modification for the BALS in \cite{Araujo2021JSAC_CE_PARAFAC}]}: In this baseline, we make a simple modification based on the MAP optimization in this paper to further enhance the performance of the BALS algorithm in \cite{Araujo2021JSAC_CE_PARAFAC} by exploiting the prior knowledge of the cascaded channels. 
\item {\bf Baseline 4 [Selected On-off protocol \cite{Liuliang_CE2020TWC}]}: This curve illustrates the performance of the estimation algorithm in \cite{Liuliang_CE2020TWC} based on the  selected on-off channel estimation protocol shown in {\figurename~\ref{protocol:c}}. We assume that the covariances of ${\bf h}_{{\rm u},k} ={\rm diag}({\bf h}_{{\rm r},1})^{-1}{\bf h}_{{\rm r},k}$ for all $k=1,2,\cdots,8$ are available, and the LMMSE estimator in \cite[Section V]{Liuliang_CE2020TWC} is adopted. In addition, we always select user $1$ as the reference user.
\end{itemize}
Note that the proposed scheme and Baselines 3 and 4 have the same  pilot overhead, i.e., $K+N+\lceil \frac{N}{M}\rceil(K-1)$.
We focus on the evaluation of the performance of cascaded channel estimation, and use the normalized MSE (NMSE) as the evaluation metric, which is given by
\begin{equation}
{\rm{NMSE}}=\frac{\sum_{k=1}^K\sum_{m=1}^M{\mathbb E} \left[\left|{\bf h}_{{\rm I},k,m}-{\hat{\bf h}}_{{\rm I},k,m}\right|_2^2\right]}
{\sum_{k=1}^K\sum_{m=1}^M {\mathbb E} \left[\left|{\bf h}_{{\rm I},k,m}\right|_2^2\right]}
.
\end{equation}
In addition, based on the proposed protocol, the BS-user channel estimation can be independent of the cascaded channel estimation by applying the signal pre-processing as shown in Section \ref{proposed_protocol}. This signal pre-processing operation will provide a theoretical $3$ dB gain for the BS-user channel estimation compared to the conventional solution, which shuts down the IRS to estimate the BS-user channel  (see Appendix \ref{app_Estimate_Hd}), and thus we do not compare the estimation performance for the BS-user channel in the simulations.

\begin{figure}
[!t]
  \centering
  \subfigure[$P_{\rm T}$ vs NMSE]{
    \label{nmse_vs_PT:a} 
    \includegraphics[width=1\columnwidth]{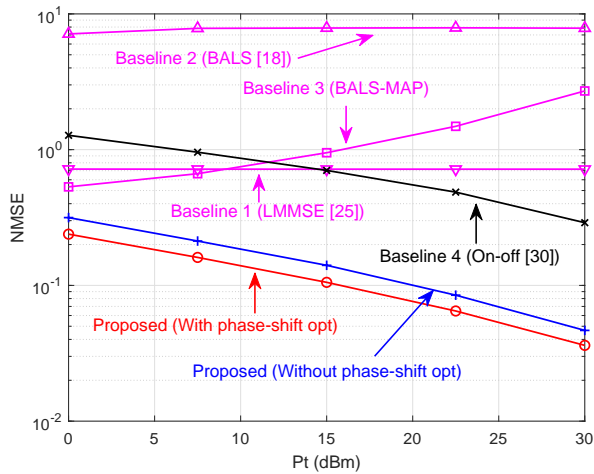}}
  \subfigure[Convergence behavior when $P_{\rm T}=15$ dBm]{
    \label{nmse_vs_PT:b} 
    \includegraphics[width=1\columnwidth]{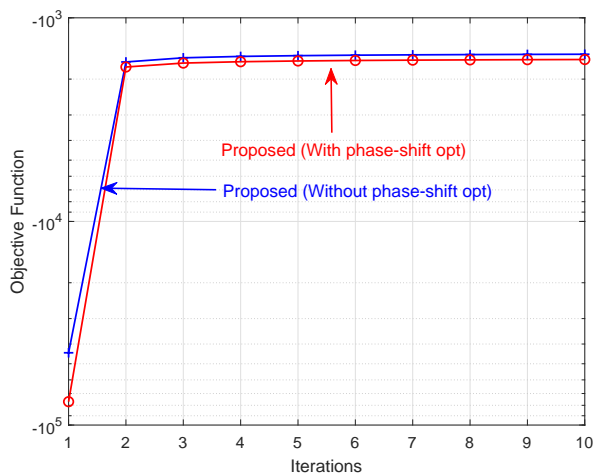}}
  \caption{The NMSE versus transmit power when $M=8$ and $N=32$.}
  \label{nmse_vs_PT} 
\end{figure}

\subsection{Simulation Results}
{\figurename~\ref{nmse_vs_PT:a}} illustrates the NMSE of different schemes with respect to the transmit power of users, in which the BS adopts a $4\times2$ uniform planar array (UPA), and the IRS adopts an $8\times4$ UPA. Thus, we have $M=8$ and $N=32$.
The BALS algorithm in \cite{Araujo2021JSAC_CE_PARAFAC} achieves the worst performance since it does not exploit the channel prior knowledge.
The performance of the LMMSE  using the traditional protocol in \cite{Kundu2021OJCSLMMSE_DFTGOOD} and \cite{Alwazani2020OJCSLMMSE_DFT} does not vary with the increase of $P_{\rm T}$ since the main bottleneck is that the number of training timeslots is smaller than $N$.
Moreover, the performance of BALS-MAP is better than that of LMMSE at a low SNR, but worsens as  $P_{\rm T}$ increases since it will reduce to BALS when $P_{\rm T}$ is infinite.
Based on the above observations, we can draw a conclusion that the traditional channel estimation protocol shown in {\figurename~\ref{protocol:b}} is not effective for exploiting the common-link structure, and thus we do not consider  Baselines 2 and 3 in the remaining simulations.
On the other hand, it is seen that the proposed protocol with the optimization-based channel estimation algorithm achieves significant gain compared to all the baselines.
In addition, the phase shifting configuration by solving ${\mathcal{P}}{(\text{B})} $ achieves a more than 3 dB gain by steering the reflected signals in Stage I to the direction with a higher SNR compared to the random configuration baseline. 
Next, in {\figurename~\ref{nmse_vs_PT:b}}, we fix the transmit power $P_{\rm T}$ at 15 dBm and show the convergence behaviors of the proposed Algorithm \ref{alg:P1} for ${\mathcal{P}}{(\text{A})}$.
One can see that the proposed algorithm converges quickly.
Note that although the solution without phase-shift optimization achieves a higher objective value, this does not imply it will have better performance since the objective functions of the two curves are different due to them adopting different ${\bm{\vartheta}}$ for the training phase shifting configuration.

\begin{figure}
[!t]
\centering
\includegraphics[width=1\columnwidth]{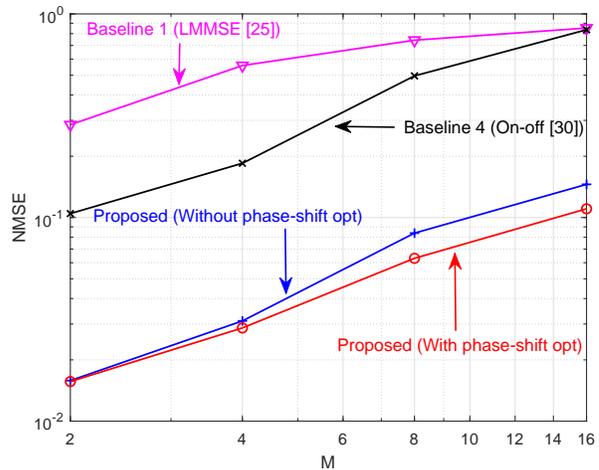}
\caption{NMSE versus $M$, when $P_{\rm T}=20$ dBm.}
\label{M_vs_NMSE}
\end{figure}

In {\figurename~\ref{M_vs_NMSE}}, we simulate the performance of different BS antenna numbers $M$ when the BS adopts the uniform linear array and the IRS is still $8\times4$ UPA. It is seen that the NMSE of all curves increases as $M$ increases since the ratio of the channel unknowns to the training observations decreases as $M$ increases. Moreover, the performance gain achieved by the phase shifting configuration increases as $M$ increases. This is because when $M$ increases, the number of training timeslots in Stage I of the proposed protocol decreases, and the probability that the random configuration scheme steers to the highest SNR direction becomes lower.

\begin{figure}
[!t]
\centering
\includegraphics[width=1\columnwidth]{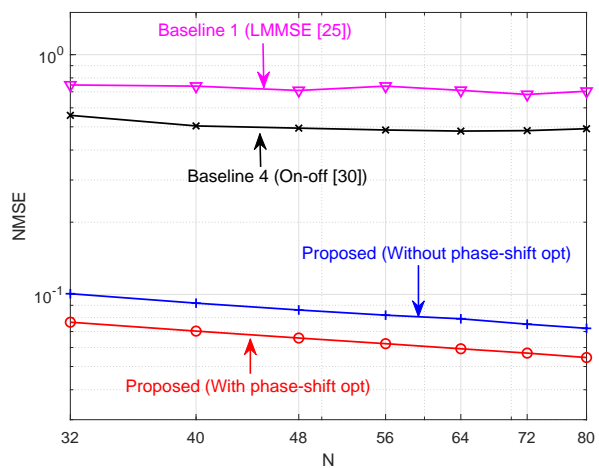}
\caption{NMSE versus $N$, when $P_{\rm T}=20$ dBm.}
\label{N_vs_NMSE}
\end{figure}

\begin{figure}
[!t]
\centering
\includegraphics[width=1\columnwidth]{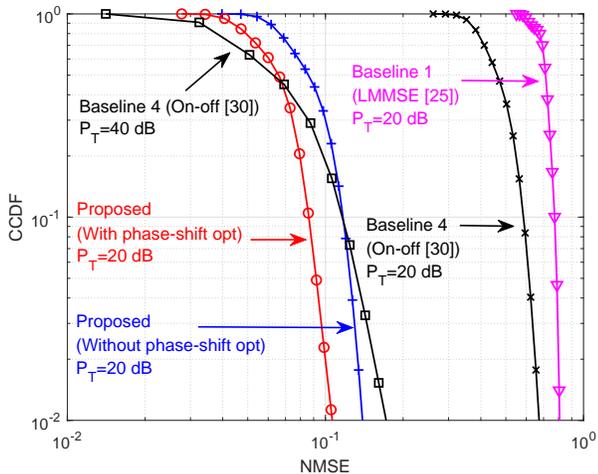}
\caption{The CCDFs for random user locations.}
\label{CCDF_location}
\end{figure}

In {\figurename~\ref{N_vs_NMSE}}, we simulate the NMSE of different schemes for different IRS sizes $N$. The BS is  $4\times2$ UPA, and the IRS is $N_1 \times 8$ UPA in which $N_1$ increases from $4$ to $10$. Note that as $N$ increases, the pilot overhead increases according to the proposed protocol but the ratio of the channel unknowns to the training observations is almost fixed.
It is seen that  the NMSE of Baselines 1 and 4 varies only a little, while the NMSE of the proposed scheme decreases as $N$ increases. This is because the channel becomes more correlated as $N$ becomes large, and the proposed scheme has a better capability of exploiting the channel prior knowledge.

Finally, we investigate the impact of user locations on the estimation performance.
In particular, we fix $N=8 \times 4$ and $M=4 \times 2$, and generate $100$ snapshots for random user locations. For each snapshot, we further generate $1000$ channel realizations with independent small-scale fading to reduce the impact of other system parameters.
{\figurename~\ref{CCDF_location}} plots the complementary cumulative distribution functions (CCDFs) of the NMSE for different snapshots.
One can see that the performance gains of the proposed scheme are irrespective of user locations.
In addition, we further increase $P_{\rm T}=40$ dBm for Baseline 4 (i.e., the selected-on-off-protocol-based scheme \cite{Liuliang_CE2020TWC}) such that it achieves a similar average NMSE to the proposed scheme with $P_{\rm T}=20$ dBm. However, Baseline 4 achieves a much worse outage performance.
This is because the performance of the selected-on-off-protocol-based scheme \cite{Liuliang_CE2020TWC} highly depends on the channel quality of the reference user, while the proposed scheme is much more robust since it does not require selecting one reference user.

\section{Conclusion}\label{conclusion}
In this paper, we proposed a novel always-ON channel estimation protocol for uplink cascaded channel
estimation in IRS-assisted MU-MISO systems.
In contrast to the existing schemes, the pilot overhead required by the proposed protocol is greatly reduced
by exploiting the common-link structure.
Based on the protocol, we formulated an optimization-based joint channel estimation problem that utilizes the combined
statistical information of the cascaded channels, and then we proposed an alternating optimization algorithm to solve the problem with the local optimum solution.
In addition, we optimized the phase shifting configuration in the proposed protocol, which may further enhance the channel estimation performance.
The simulation results demonstrated that the proposed protocol using the optimization based joint channel estimation algorithm achieves a more than $15$ dB gain compared to the benchmark. In addition, the proposed optimized phase shifting configuration  achieves a more than $3$ dB gain compared to the random configuration scheme.

\appendices
\section{Proof of Lemma \ref{lemma0}}\label{proof_lemma0}
Define the virtual reference channel by ${\bf H}_{\rm v}={\bf G}^{\rm T} {\rm diag} ({\bf H}_{\rm r}\bar{\bf x})$. Based on \eqref{equ:rbar}, we have
\begin{equation}\label{equ:app0_Hv}
\begin{aligned}[b]
\left[\bar{\bf r}_1,\cdots,\bar{\bf r}_N \right]
&={\bf H}_{\rm v}  {\bm \Phi}
.
\end{aligned}
\end{equation}
Thus ${\bf H}_{\rm v}$ can be perfectly estimated by:
\begin{equation}\label{equ:app0_Hv2}
\begin{aligned}[b]
{\bf H}_{\rm v}= \left[\bar{\bf r}_1,\cdots,\bar{\bf r}_N \right]
{\bm \Phi}^{-1} .
\end{aligned}
\end{equation}
We further define the $K$ relative channels by ${\bf h}_{{\rm A},k}={\rm diag} ({\bf H}_{\rm r}\bar{\bf x})^{-1} {\bf h}_{{\rm r},k}$, and ${\bf H}_{\rm A}=[{\bf h}_{{\rm A},1},{\bf h}_{{\rm A},2},\cdots,{\bf h}_{{\rm A},K}]$.
Then the cascaded channels become ${\bf H}_{{\rm I},k}= {\bf H}_{\rm v}{\rm diag}({\bf h}_{{\rm A},k})$. Therefore, the remaining task is to perfectly estimate ${\bf H}_{\rm A}$.

Based on the assumption on $\bf G$ and ${\bf H}_{\rm r}$, we have
${\bf H}_{\rm v}={\bf F}_{\rm B} {\ddot{\bf G}}^{\rm T} {\bf F}_{\rm R}^{\rm T} {\rm diag} ({\bf F}_{\rm R} {\ddot{\bf H}}_{\rm r}\bar{\bf x})$ where ${\ddot{\bf H}}_{\rm r}=[{\ddot{\bf h}}_{{\rm r},1},\cdots,{\ddot{\bf h}}_{{\rm r},K}]$.
Define ${\bf V}=[{\bf v}_{1},{\bf v}_{2},\cdots,{\bf v}_{M}]$ which is given by
\begin{equation}\label{equ:app0_V}
\begin{aligned}[b]
{\bf v}_m={\rm diag} ({\bf F}_{\rm R} {\ddot{\bf H}}_{\rm r}\bar{\bf x}){\bf F}_{\rm R} {\ddot{\bf g}}_m,
\end{aligned}
\end{equation}
and thus ${\bf H}_{\rm v}={\bf F}_{\rm B} {\bf V}^{\rm T}$.
Using the independence of ${\ddot{\bf g}}_m$ and ${\ddot{\bf h}}_{{\rm r},k}$, we have ${\mathbb E}[{\bf v}_i {\bf v}_j^{\rm H}]={\bm 0}$. Since all ${\bf v}_{m}$ ($m=1,2,\cdots,M$) follow joint multivariate normal distribution, ${\bf v}_{m}$ for all $m$ are pairwise  independent to each other.
In addition, since ${\bf C}_m^{(k)}={\mathbb E}\left[ {\bf h}_{{\rm I},k,m} {\bf h}_{{\rm I},k,m}^{\rm H} \right]$ is full-rank for all $k$ and $m$, ${\mathbb E}[{\bf v}_m {\bf v}_m^{\rm H}]$ is also full-rank for all $m$ with properly-designed $\bar{\bf x}$.

Next, using ${\bf R}_{\ell}$ in \eqref{equ:R1} and \eqref{equ:R2L1}, we have
\begin{equation}\label{equ:app0_RL}
\begin{aligned}[b]
\tilde{\bf R}_{\ell} &= {\bf F}_{\rm B}^{-1} {\bf R}_{\ell} {\bf X}^{-1}\\
&={\bf F}_{\rm B}^{-1} {\bf G}^{\rm T} {\rm diag} ({\bm \theta}_\ell) {\bf H}_{\rm r}\\
&={\bf F}_{\rm B}^{-1} {\bf H}_{\rm v} {\rm diag} ({\bm \theta}_\ell) {\bf H}_{\rm A}\\
&={\bf V}^{\rm T} {\rm diag} ({\bm \theta}_\ell) {\bf H}_{\rm A}
.
\end{aligned}
\end{equation}
Stacking all $\tilde{\bf R}_{\ell}$, we have
\begin{equation}\label{equ:app0_mesure}
\begin{aligned}[b]
\begin{bmatrix}\tilde{\bf R}_1\\ \vdots \\ \tilde{\bf R}_{L_1}\end{bmatrix}
=\begin{bmatrix}{\bf V}^{\rm T}{\rm diag} ({\bm \theta}_1)\\ \vdots \\ {\bf V}^{\rm T}{\rm diag} ({\bm \theta}_{L_1})\end{bmatrix}
{\bf H}_{\rm A}.
\end{aligned}
\end{equation}
Define ${\bm \Psi}=[{\rm diag} ({\bm \theta}_1){\bf V},\cdots,{\rm diag} ({\bm \theta}_{L_1}){\bf V}]^{\rm T}$. Now we need to prove ${\text{rank}}({\bm \Psi})={\rm{min}}\{N,ML_1\}$ with probability one, and the whole proof is completed.

By permuting the columns of ${\bm \Psi}$, we have a new matrix
$\bar{\bm \Psi}=[{\rm diag} ({\bf v}_1)\bar{\bm \Phi},\cdots,{\rm diag} ({\bf v}_M)\bar{\bm \Phi}]^{\rm T}$
where $\bar{\bm \Phi}=[{\bm \theta}_1,\cdots,{\bm \theta}_{L_1}]$.
Then, it is equivalent to prove that ${\text{rank}}(\bar{\bm \Psi})={\rm{min}}\{N,ML_1\}$ with probability one. We prove it by induction.
Define $\bar{\bm \Psi}_m=[{\rm diag} ({\bf v}_1)\bar{\bm \Phi},\cdots,{\rm diag} ({\bf v}_m)\bar{\bm \Phi}]^{\rm T}$.
Since $\bar{\bm \Phi}$ is semi-orthogonal, ${\text{rank}}(\bar{\bm \Psi}_1)=L_1$ with probability one.
Let ${\text{rank}}(\bar{\bm \Psi}_{m-1})={\rm{min}}\{N,(m-1)L_1\}$ with probability one, and the rest task is to prove ${\text{rank}}(\bar{\bm \Psi}_m)={\rm{min}}\{N,mL_1\}$ with probability one.
We prove it by contradiction.
Consider the case when ${\text{rank}}(\bar{\bm \Psi}_{m-1})=(m-1)L_1$ which is smaller than $N$ but ${\text{rank}}(\bar{\bm \Psi}_m)<{\rm{min}}\{N,mL_1\}$.
Since ${\rm diag} ({\bf v}_m) {\bm \theta}_i$ and ${\rm diag} ({\bf v}_m) {\bm \theta}_j$ are orthogonal for $i \neq j$, there shall exists $\bf x$ which satisfies:
\begin{equation}\label{equ:app0_x}
\begin{aligned}[b]
[{\rm diag} ({\bf v}_1)\bar{\bm \Phi},\cdots,{\rm diag} ({\bf v}_{m-1})\bar{\bm \Phi}]{\bf x}={\rm diag} ({\bf v}_m) {\bm \theta}_\ell,
\end{aligned}
\end{equation}
for some $1\leq \ell \leq L_1$ such that ${\text{rank}}(\bar{\bm \Psi}_m)<{\rm{min}}\{N,mL_1\}$ is true.
However, since ${\bf v}_m$ is independent to ${\bf v}_1,{\bf v}_2,\cdots,{\bf v}_{m-1}$ with full-rank covariance matrix, equation \eqref{equ:app0_x} is inconsistent (which has no solution) with probability one, and the whole proof is finished.

\section{Estimation on the BS-User Channel}\label{app_Estimate_Hd}
Based on ${\bm \theta}_0=-{\bm \theta}_1$, we have
\begin{equation}\label{equ:r_direct}
\begin{aligned}[b]
{\bf R}_{0} &= \frac{1}{2}\left({{\bf Y}}_0+{{\bf Y}}_1\right) {\bf X}^{-1}\\
&={\bf H}_{\rm d}+\tilde{\bf Z}_0
,
\end{aligned}
\end{equation}
where $\tilde{\bf Z}_0$ is the noise matrix consisting of $MK$ i.i.d. complex Gaussian variables following ${\cal{CN}}({ 0},\frac{1}{2K}\sigma_0^2) $.
Let ${\bf r}_{{0},k}$ be the $k$-th column of ${\bf R}_{0}$.
The  BS-user direct channel for the $k$-th user can be estimated by the
LMMSE
estimator \cite{Kay1993statisticalSP}:
\begin{equation}\label{equ:hat_direct}
\begin{aligned}[b]
\hat{\bf h}_{{\rm d},k}={\bf C}_{{\rm d},k} \left({\bf C}_{{\rm d},k}+ \frac{\sigma_0^2}{2K} {\bf I}_M\right)^{-1} {\bf r}_{0,k},
\end{aligned}
\end{equation}
where ${\bf C}_{{\rm d},k}={\mathbb E}\left[ {\bf h}_{{\rm d},k} {\bf h}_{{\rm d},k}^{\rm H} \right]$
is the covariance matrix of the direct channel from the BS to the $k$-th user.
Note that, as shown in \eqref{equ:r_direct} and \eqref{equ:hat_direct}, the proposed protocol may achieve a $3$ dB performance gain compared to the existing works, \cite{Kundu2021OJCSLMMSE_DFTGOOD,Alwazani2020OJCSLMMSE_DFT,Liuliang_CE2020TWC}, on the estimation of the BS-user channels since it exploits doubled observation samples.

\section{Proof of Lemma \ref{eq_decomposite}}\label{proof_lemma1}
The objective function of ${\mathcal{P}}{(\text{A})}$ can be denoted by the function of the cascaded channel coefficients $\{{\bf h}_{{\rm I},k,m}\}$, as follows:
\begin{equation}\label{equ:obj_f_A2}
\begin{aligned}[b]
&f_{{\rm A}}({\bf H}_{\rm g}, {\bf H}_{\rm u})
=f_{{\rm A}}(\{{\bf h}_{{\rm I},k,m}\})\\
&=-\sum_{\ell=1}^{L_1} \sum_{k=1}^K
{\frac{1}{\sigma_{\ell}^{2}} \left\|
\tilde{{\bf r}}_{\ell,k}- \sum_{m=1}^M {\bf h}_{{\rm I},k,m}^{\rm T} {\bm \theta}_{\ell}
\right\|^2 }\\
&\quad-\sum_{\ell=L_1+1}^{N}
{ \frac{1}{\bar\sigma_{\ell}^{2}} \left\|
\bar{\bf r}_{\ell}-\sum_{k=1}^K \sum_{m=1}^M \bar{x}_k {\bf h}_{{\rm I},k,m}^{\rm T} {\bm \theta}_{\ell}
\right\|^2 }
\\
&\quad-\sum_{m=1}^M \sum_{k=1}^K {\bf h}_{{\rm I},k,m}^{\rm H} {{\bf C}_m^{(k)}}^{-1} {\bf h}_{{\rm I},k,m}  .
\end{aligned}
\end{equation}
Therefore, if one optimal solution $\left\{{\bf H}_{\rm g}^\star,{\bf H}_{\rm u}^\star \right\} \in {\cal A}$, any $\left\{{\bf H}_{\rm g},{\bf H}_{\rm u}\right\}$ pair in set ${\cal A}$ is an optimal solution of ${\mathcal{P}}{(\text{A})}$, and the lemma is proved.

\section{Proof of Lemma \ref{fu_convex}}\label{proof_lemmafu}
Denote $\ddot{\bf h}_{\rm u}={\rm{vec}}({\bf H}_{\rm u})$. The objective function $f_{{\rm u}}$ in \eqref{equ:obj_f_hrk} is given by
\begin{equation}\label{equ:proof_fu_1}
\begin{aligned}[b]
&f_{{\rm u}}(\ddot{\bf h}_{\rm u})
=\sum_{\ell=1}^{L_1}
{\frac{1}{\sigma_{\ell}^{2}} \left\|
{\rm{vec}} (\tilde{{\bf R}}_{\ell})- \left({\bf I}_K \otimes {\bf D}_{\ell}\right) \ddot{\bf h}_{\rm u}
\right\|^2_{2} }\\
&\quad+\sum_{\ell=L_1+1}^{N}
{ \frac{1}{\bar\sigma_{\ell}^{2}} \left\|
\bar{\bf r}_{\ell}-\left({\bar{\bf x}}^{\rm T} \otimes {\bf D}_{\ell}\right) \ddot{\bf h}_{\rm u}
\right\|^2_2 }
+ \ddot{\bf h}_{\rm u}^{\rm H} {\bf C}_{{\rm u}} \ddot{\bf h}_{\rm u}
,
\end{aligned}
\end{equation}
where ${\bf C}_{{\rm u}}={\rm{blkdiag}}({\bf C}_{{\rm u},1},{\bf C}_{{\rm u},2},\cdots,{\bf C}_{{\rm u},K})$.
The second order derivative of $f_{{\rm u}}(\ddot{\bf h}_{\rm u})$ is given by
\begin{equation}\label{equ:proof_fu_2}
\begin{aligned}[b]
\frac{ \partial^2 f_{{\rm u}}(\ddot{\bf h}_{\rm u})}{\partial \ddot{\bf h}_{\rm u} \partial \ddot{\bf h}_{\rm u}^{\rm H}}
&=2{\bf C}_{{\rm u}}
+2{\bf I}_K \otimes \left(\sum_{\ell=1}^{L_1} \frac{1}{\sigma_{\ell}^{2}} {\bf D}_{\ell}^{\rm H} {\bf D}_{\ell}\right) \notag\\
&\quad + 2\left(\bar{\bf x}^\ast \bar{\bf x}^{\rm T}\right)\otimes \left(\sum_{\ell=L_1+1}^{N} \frac{1}{\bar\sigma_{\ell}^{2}} |\bar{x}_k|^2 {\bf D}_{\ell}^{\rm H} {\bf D}_{\ell}\right)
,
\end{aligned}
\end{equation}
which is a  Hermitian positive semi-definite matrix. Thus, the lemma is proved.

\section{Proof of Lemma \ref{fg_convex}}\label{proof_lemmafg}
The second order derivative of $f_{{\rm g},m}({\bf h}_{{\rm g},m})$ in \eqref{equ:obj_f_gm} is given by
\begin{equation}\label{equ:proof_fg_1}
\begin{aligned}[b]
&\frac{ \partial^2 f_{{\rm g},m}({\bf h}_{{\rm g},m})}{\partial {\bf h}_{{\rm g},m} \partial {\bf h}_{{\rm g},m}^{\rm H}}
=2{\bf C}_{{\rm g},m}^{-1}
+2\sum_{k=1}^K\sum_{\ell=1}^{L_1} \frac{1}{\sigma_{\ell}^{2}} {\bf b}_{\ell,k}^\ast {\bf b}_{\ell,k}^{\rm T}\\
&\quad+ 2\sum_{\ell=L_1+1}^{N}\frac{1}{\bar\sigma_{\ell}^{2}}
\left(\sum_{k =1}^K \bar{x}_k {\bf b}_{\ell,k}^{\rm T}\right)^{\rm H}
\left(\sum_{k =1}^K \bar{x}_k {\bf b}_{\ell,k}^{\rm T}\right)
,
\end{aligned}
\end{equation}
which is a  Hermitian positive semi-definite matrix. Thus, $f_{{\rm g},m}({\bf h}_{{\rm g},m})$ is a convex quadratic function of ${\bf h}_{{\rm g},m}$. Then, the objective function $\sum_{m=1}^M f_{{\rm g},m}({\bf h}_{{\rm g},m})$ is a convex quadratic function of $\{{\bf h}_{{\rm g},1},{\bf h}_{{\rm g},2},\cdots,{\bf h}_{{\rm g},M}\}$.

\bibliographystyle{IEEEtran}
\bibliography{IEEEabrv,mybib_draft2}

\end{document}